\documentclass[a4paper]{article}
\usepackage[margin=1in]{geometry} % full-width

\usepackage{amsmath}
\usepackage{amsthm}
\usepackage{amssymb}

\usepackage[utf8]{inputenc}
\usepackage{hyperref}
\hypersetup{
	unicode,
	pdfauthor={Feyza Başpehlivan, Burak Dagli, Osman Emre Delialioglu, Saleh Sultansoy},
	pdftitle={Why should we search for vector-like leptons},
	pdfsubject={High Energy Physics - Phenomenology},
	pdfkeywords={article, template, simple},
	pdfproducer={LaTeX},
	pdfcreator={pdflatex}
}
\DeclareUnicodeCharacter{2212}{-}
\DeclareUnicodeCharacter{207B}{+}
\DeclareUnicodeCharacter{2009}{\,}
\usepackage{graphicx}
\usepackage{float}
\usepackage{lipsum}
\usepackage{appendix}
\usepackage{multirow}

\usepackage[sorting=none]{biblatex}
\title{Why should we search for vector-like leptons?}
\author{Feyza Baspehlivan$^1$, Burak Dagli${^{1*}}$, Osman Emre Delialioglu$^1$, Saleh Sultansoy$^{1,2}$}

\date{
	\small{$^1$TOBB University of Economics and Technology, Ankara, Turkey \\ %\texttt{\{auth1, auth3\}@org1.edu}\\%
	$^2$ANAS Institute of Physics, Baku, Azerbaijan \\% \texttt{auth3@inst2.edu}\\[2ex]%
}
}

\begin{document}
	\maketitle
    \vspace{5mm}
	\begin{abstract}
\noindent
There are two strong arguments in favor of vector-like leptons and quarks: Flavor Democracy call for them, and $E_{6}$ GUT predicts existence of iso-singlet quarks and iso-doublet leptons. Vector-like quarks (VLQ) are extensively searched by ATLAS and CMS collaborations, but this is not the case for vector-like leptons (VLL), while they have actually similar status from phenomenology viewpoint: a few papers published by ATLAS and CMS are mostly devoted to search for vector-like partners of the third SM family leptons. In this study we argue that a search for vector-like leptons related to first and second SM families should be included into the new physics search programs of energy-frontier colliders.\\

\noindent
We consider production of vector-like partners of the first SM family leptons at the HL-LHC, HE-LHC, FCC, ILC, CLIC, Muon Collider, as well as, at $ep$ and $\mu$p colliders. As for decays of vector-like leptons, we present branching ratios formulas to different channels for the most general case. Since there are many different production and decay channels for charged and neutral vector-like leptons, relevant studies should be done systematically. We invite the High Energy Physics community (both experimenters and phenomenologists) to actively participate in research on this topic.
		
	\end{abstract}
\mbox{}
\vfill
* Corresponding author:  \href{mailto:dburakdagli@gmail.com}{dburakdagli@gmail.com} , \href{mailto:burakdagli@etu.edu.tr}{burakdagli@etu.edu.tr}

\newpage
\tableofcontents

	\newpage
	\section{Introduction}
	\label{sec:intro}

The Standard Model (SM) describes well a huge number of experimental results in particle physics \cite{RevParPhy}. With the discovery of the Higgs boson by ATLAS and CMS collaborations \cite{ObsNewPar,ObsNewBos} electroweak sector of the SM have been completed, whereas this is not the case for QCD sector: confinement hypothesis should be clarified, and quark-hadron transition should be understood.\\
On the other hand, there are a lot of particle physics related topics which have not been resolved by the SM. A few examples are: left-right asymmetry is put by hand (left-handed components of fermions are SU(2) doublets and right-handed components are singlets), mass and mixing patterns of the SM fermions seems to be accidental (for mass values of the SM fermions see Table 1), huge difference between electroweak and Planck scales etc. (for a more complete list see i.e. reviews \cite{Sultansoy:1997hf,Sultansoy:1997hg}).

\renewcommand{\arraystretch}{1.5}
	\begin{table}[H]
		\centering
		\caption{Mass pattern of charged leptons and quarks \cite{RevParPhy}.}
		\begin{tabular}{|c|c|c|c|}
			\hline
			\textbf {} & \textbf{charged leptons} & \textbf{Up type quarks} & \textbf{Down type quarks}  \\
			\hline \hline
			$1^{st}$ Family & 0.5109989461 $\pm$ 0.0000000031  MeV & $2.16^{+0.49}_{-0.26}$ MeV & $4.67^{+0.48}_{-0.17}$ MeV \\
			\hline
			$2^{nd}$ Family & 105.6583745 ± 0.0000024 MeV  & 1.27 $\pm$ 0.02 GeV & $93^{+11}_{-5}$ MeV \\
			\hline
			$3^{rd}$ Family & 1776.86 ± 0.12 MeV & 172.76 $\pm$ 0.30 GeV & $4.18^{+0.03}_{-0.02}$ GeV \\
			\hline
		\end{tabular}
		\label{tbl:1}
	\end{table}
\noindent	
The Flavor Democracy (FD) hypothesis (see review \cite{Sultansoy:2007fd} and references therein) may enlighten mass pattern of the SM fermions: heavy t-quark (with mass of order of vacuum expectation value of Higgs field) and much lighter other SM fermions. In this aspect inclusion of new leptons and quarks (the fourths SM family, vector-like leptons (VLL) and quarks (VLQ)) is necessary. The fourth SM family is almost excluded by experimental data on Higgs boson production and decays. Therefore, introduction of VLLs and VLQs remains as the sole option. It should be mentioned that existence of new iso-singlet down-type quarks and vector-like lepton doublets (per each SM family) is predicted by $E_{6}$ Grand Unified Theory (GUT) \cite{Gursey:1975ki,Gursey:1981kf} (see also review \cite{Hewett:1988xc}).\\\\
It should be noted that from particle phenomenology viewpoint vector-like leptons has the same status as vector-like quarks. A search for VLQs is widely performed by ATLAS and CMS collaborations (see for example \cite{PhysRevLett.109.071801,Aad:2015sf,PhysRevLett.121.211801,Sirunyan:2017ss,Sirunyan:2019sfs,Sirunyan:2019sfp}), whereas this is not the case for VLLs. A severe phenomenology studies related to search for VLLs at the LHC are concentrated on the vector-like partners of the third SM family leptons \cite{Kumar:2015vl,Bhattiprolu:2019pf}. The reason of this is based on the predominance of mixtures in the quark sector (CKM matrix) among neighboring families \cite{RevParPhy}:

\begin{equation}
V_{CKM} = \begin{pmatrix}
0.97401\pm0.00011 & 0.22650\pm0.00048 & 0.00361^{+0.00011}_{-0.00009} \\
0.22636\pm0.00048 & 0.97320\pm0.00011 & 0.04053^{+0.00083}_{-0.00061} \\
0.00854^{+0.00023}_{-0.00016} & 0.03978^{+0.00082}_{-0.00060} & 0.999172^{+0.000024}_{-0.000035} \\
\end{pmatrix}
\end{equation}
As for the new quarks, this assumption can be considered natural for the fourth SM family quarks (the dominance of the fourth family quarks mixing with the third family quarks). Concerning for vector-like quarks, there is not any serious reason for their mixture with the third family to be dominant. On the other hand, the lepton mixing matrix (PMNS matrix) is quite different from the CKM matrix and mixings of adjacent families are not dominant \cite{esteban2020fate}:
\begin{equation}
|U|^{with SK-atm}_{3\sigma} = \begin{pmatrix}
0.801\rightarrow0.845 & 0.513\rightarrow0.579 & 0.143\rightarrow0.155 \\
0.234\rightarrow0.500 & 0.471\rightarrow0.689 & 0.637\rightarrow0.776 \\
0.271\rightarrow0.525 & 0.477\rightarrow0.694 & 0.613\rightarrow0.756 \\
\end{pmatrix}
\end{equation}
Keeping in mind experimental results on mixings in lepton sector, which are drastically different comparing to CKM of quark sector, assumption that new vector-like leptons are dominantly mixed with third SM family leptons is not sufficiently motivated. It is quite possible that new VLLs are mixed dominantly with first or second SM family leptons (in the framework $E_{6}$ based models it is quite possible that interfamily mixings between new fermions and the SM fermions are dominant). This possibility has been considered in a few papers: associated production of neutral and charged vector-like leptons at the LHC is analyzed in \cite{_zcan_2009}, and pair production of charged vector-like leptons at the LHC and FCC is recently considered in \cite{acar2021search}.\\\\
In addition, vector-like leptons can offer solutions to some problems of particle physics. For example, they provide solution for “Cabibbo Angle Anomaly” and improve electroweak fit comparing to SM \cite{Crivellin:2020ge}. Then, they can help getting a realistic Higgs mass without problematic light top partners in the framework of Composite Higgs models \cite{Carmona:2015an,PhysRevLett.116.251801,Carmona:2017rb}. In ref \cite{guedes2021new} current and future hadron collider limits on new vector-like leptons with exotic decays, namely into a Standard Model charged lepton and a stable particle like a dark photon, are considered and the interplay between the dark photon and the vector-like lepton in generating the observed dark matter relic abundance and the complementarity of collider searches and dark matter phenomenology is discussed. Vector-like leptons together with additional scalars explain discrepancies between measured values of the electron and muon anomalous magnetic moments and SM predictions \cite{Hiller:2020am}. They also provide templates of asymptotically safe SM extensions with physical Higgs, top, and bottom masses, and which connect the relevant SM and BSM couplings at TeV energies with an interacting fixed point at highest energies \cite{PhysRevD.102.095023}. Multi-lepton signatures of vector-like leptons of this model at the LHC have been considered in \cite{Bismann:2021ml}. Finally, quasi-stable charged vector-like leptons may provide opportunity to realize VLL-catalyzed fusion by analogy with muon-catalyzed fusion (see, for example \cite{doi:10.1063/1.5135483}).\\\\
This paper is devoted to general consideration of production and decays of the first SM family related iso-singlet and iso-doublet vector-like leptons. In the next section, we present motivations for existence of iso-singlet down-type quarks and vector-like leptons based on FD hypothesis and E6 GUT. Lagrangians for interactions of vector-like leptons with intermediate vector bosons are given in section 3. Production cross-sections for iso-singlet and iso-doublet VLLs at the LHC and future energy frontier colliders are considered in section 4. Section 5 is devoted to consideration of possible decay modes of new leptons. Several example processes are considered in section 6. Finally, in section 7 we present our conclusions and recommendations.
	\section{Phenomenological arguments favoring vector-like leptons and quarks}
There are two strong arguments for existence of VLLs and VLQs: Flavor Democracy (down-up) and $E_{6}$ GUT (up-down). Below we briefly discuss these arguments.
	    \subsection{Flavor Democracy calls for iso-singlet down quarks and vector-like leptons}
	    \label{sec:2.1}
As mentioned in introduction, ass pattern of the SM fermions may be enlightened by the flavor democracy (democratic mass matrix) hypothesis through inclusion of vector-like leptons and quarks. Recently this possibility is reconsidered in \cite{acar2021search,Kaya:2018rsr}. In this section, we briefly summarize corresponding part of \cite{Kaya:2018rsr}.\\

\noindent
\textbf{a) Iso-singlet quark}\\

\noindent
The quark sector of SM is modified by addition of iso-singlet down-type quark D:
    \begin{equation}
	\binom{u_L}{d_L},\binom{c_L}{s_L},\binom{t_L}{b_L},u_R,d_R,c_R,s_R,t_R,b_R,D_L,D_R
	\end{equation}
\noindent
Let us note that it is unnatural for up-type quarks to have vector-like partners under the flavor democracy hypothesis, because of high mass of t-quark ($m_{t}\simeq175$ GeV, which is close to vacuum expectation value of the Higgs field), whereas the addition of the vector-like partners of the down-type quarks explains why the b-quark has a much lower mass than the t-quark.
In the case of full Flavor Democracy, the mass matrix of the up-type quarks can be written as\\
\begin{equation}
\centering
\begin{tabular}{cccc}
 & $u_R$ & $c_R$ & $t_R$ \\
$u_L$ & $a\eta$ &  $a\eta$ &  $a\eta$  \\
$c_L$ & $a\eta$ &  $a\eta$ &  $a\eta$ \\
$t_L$ & $a\eta$ &   $a\eta$  &  $a\eta$\\
\end{tabular}
\end{equation}
and mass matrix of down type quarks is \\
\begin{equation}
\centering
\begin{tabular}{ccccc}
 & $d_R$ & $s_R$ & $b_R$ & $D_R$ \\
$d_L$ & $a\eta$ &  $a\eta$ &  $a\eta$  &  $a\eta$ \\
$s_L$ & $a\eta$ &  $a\eta$ &  $a\eta$ &  $a\eta$ \\
$b_L$ & $a\eta$ &   $a\eta$  &  $a\eta$ &  $a\eta$ \\
$D_L$ & $M$ &   $M$  &  $M$ &  $M$ \\
\end{tabular}
\end{equation}
where $\eta = 246$ GeV is vacuum expectation value of Higgs field and M (M $>> \eta$) is the new physics scale that determines the mass of iso-singlet quark. In this case $m_{u} = m_{c} = 0$ and $m_{t} = 3a\eta$ for up type quarks, $m_{d} = m_{s} = m_{b} = 0$ and $m_{D} = 3a\eta + M = m_{t} + M$ for down type quarks.\\\\
Keeping in mind that the mass values of first family quarks are small, we consider mass generation for b, c and s quarks by moderate modifications of full democracy. With the following modification of mass matrix of up quarks:\\
\begin{equation}
\centering
\begin{tabular}{cccc}
 & $u_R$ & $c_R$ & $t_R$ \\
$u_L$ & $a\eta$ &  $a\eta$ &  $a\eta$  \\
$c_L$ & $a\eta$ &  $a\eta$ &  $a\eta$ \\
$t_L$ & $a\eta$ &   $a\eta$  &  $(1+\alpha_{c})a\eta$\\
\end{tabular}
\end{equation}
we obtain $m_{t} = 3a\eta$, $m_{c} = 2\alpha_{c}m_{t}/9$ and $\mu = 0$. Therefore, $a = m_{t}/3\eta = 0.233$ and $\alpha_{c} = 9m_{c}/2m_{t} = 3.3 \times 10^{−2}$.\\\\
In order to generate masses for b and s quarks, we consider following modification of mass matrix for down quarks:\\
\begin{equation}
\centering
\begin{tabular}{ccccc}
 & $d_R$ & $s_R$ & $b_R$ & $D_R$ \\
$d_L$ & $a\eta$ &  $a\eta$ &  $a\eta$  &  $(1-\alpha_b)a\eta$ \\
$s_L$ & $a\eta$ &  $a\eta$ &  $a\eta$ &   $(1-\alpha_b)a\eta$ \\
$b_L$ & $a\eta$ &   $a\eta$  &  $(1+\alpha_{s})a\eta$ &   $(1-\alpha_b)a\eta$ \\
$D_L$ & $(1-\beta_b)M$ &   $(1-\beta_b)M$  &  $(1-\beta_b)M$ &  $M$ \\
\end{tabular}
\end{equation}
For $M = 2000$ GeV, $\alpha_{b} = \beta_{b} = 1.32 \times 10^{−2}$ and $\alpha_{s} = 2.48 \times 10^{−4}$ we obtain $m_{D} = 2168$ GeV, $m_{b} = 4.18$ GeV, $m_{s} = 95.2$ MeV, $m_{d} = 0$.\\\\
It should be noted that D-quark provides an opportunity to explain experimental discrepancy in the first road of CKM matrix \cite{Belfatto:2021at}.\\

\noindent
\textbf{b) Vector-like leptons}\\

\noindent
In a similar manner low value of $\tau$ lepton mass can be provided by adding an iso-singlet charged
lepton:
    \begin{equation}
	\binom{e_L}{\nu_{e_L}},\binom{\mu_L}{\nu_{\mu_L}},\binom{\tau_L}{\nu_{\tau_L}},e_R,\nu_{e_R},\mu_R,\nu_{\mu_R},\tau_R,\nu_{\tau_R},E_L,E_R
	\end{equation}
Let us emphasize that right-handed neutrinos should be included into SM since (according to lepton-quark symmetry) there are counterparts of right-handed components of up quarks. This statement is confirmed by the observation of neutrino oscillations. Earlier, absence of $\nu_{R}$’s was postulated mistakenly because of V-A structure of charged weak currents.\\\\
Masses of $\tau$-lepton and muon are generated by following modification of the charged lepton mass matrix:\\
\begin{equation}
\centering
\begin{tabular}{ccccc}
 & $e_R$ & $\mu_R$ & $\tau_R$ & $E_R$ \\
$e_L$ & $a\eta$ &  $a\eta$ &  $a\eta$  &  $(1-\alpha_\tau)a\eta$ \\
$\mu_L$ & $a\eta$ &  $a\eta$ &  $a\eta$ &   $(1-\alpha_\tau)a\eta$ \\
$\tau_L$ & $a\eta$ &   $a\eta$  &  $(1+\alpha_\mu)a\eta$ &   $(1-\alpha_\tau)a\eta$ \\
$E_L$ & $(1-\beta_\tau)M$ &   $(1-\beta_\tau)M$  &  $(1-\beta_\tau)M$ &  $M$ \\
\end{tabular}
\label{eqn:7}
\end{equation}
For $M = 2000$ GeV, $\alpha_{\tau} = \beta_{\tau} = 5.58 \times 10^{-3}$ and $\alpha_{\mu} = 2.73 \times 10^{-4}$ this mass matrix leads to $m_L = 2171$ GeV, $m_{\tau} = 1.777$ GeV, $m_{\mu} = 104.7$ MeV, $m_e = 0$.\\\\
Zero masses for the SM neutrinos can be provided by introduction iso-singlet vector-like heavy neutral lepton $N_L$ and $N_R$. Corresponding mass matrix has a form:\\
\begin{equation}
\centering
\begin{tabular}{ccccc}
 & $\nu_{e_R}$ & $\nu_{\mu_R}$ & $\nu_{\tau_R}$ & $N_R$ \\
$\nu_{e_L}$ & $a\eta$ &  $a\eta$ &  $a\eta$  &  $a\eta$ \\
$\nu_{\mu_L}$ & $a\eta$ &  $a\eta$ &  $a\eta$ &   $a\eta$ \\
$\nu_{\tau_L}$ & $a\eta$ &   $a\eta$  &  $a\eta$ &   $a\eta$ \\
$N_L$ & $M$ &   $M$  &  $M$ &  $M$ \\
\end{tabular}
\label{eqn:8}
\end{equation}
Diagonalization of this matrix leads to $m(\nu_e) = m(\nu_\mu) = m(\nu_\tau) = 0$ and $m_N = M + m_t$.\\\\
In the case of iso-doublet vector-like leptons, mass matrices in Equations (9)-(10) should be replaced with transposing ones. The resulting lepton mass values is not change.
        \subsection{\texorpdfstring{$E_{6}$}{} GUT predicts iso-singlet down-type quarks and iso-doublet vector-like leptons}
The first family fermion sector of the $E_6$-induced model has the following $SU_c (3) \times SU_w (2) \times U_Y (1)$ structure: 
    \begin{equation}
	\binom{u_L}{d_L}\ u_R\ d_R\ D_L\ D_R\ \binom{\nu_{e_L}}{e_L}\ \nu_{e_R}\ e_R\ \binom{N_{e_L}}{E_L}\ \binom{N_{e_L}}{E_R}\ \mathcal{N}_e\\
	\end{equation}
Similar structure is a case for second and third family fermions. Therefore, quark sector of the SM is extended by addition of three new iso-singlet down-type quarks. As for lepton sector, there are three new charged leptons, three new neutral Dirac leptons and three neutral Majorana leptons. Hereafter we assume that interfamily mixings are dominant which means that new leptons will decay into their SM partners.\\\\
A search for $E_6$ iso-singlet quarks at the LHC was proposed in \cite{Mehdiyev:2007ps,Sultansoy:978632,Sultansoy:2008te,Mehdiyev:2008dt} fifteen years ago.
    \section{Interaction Lagrangians}
Lagrangians for interactions of leptons with intermediate vector-bosons are given below:

\begin{equation}
\begin{split}
L_{int}^{SM}
& = eA_{\mu}\overline{E}\gamma^{\mu}E - \frac{g}{2\sqrt{2}}[W_{\mu}^{-}\overline{e}\gamma^{\mu}(1-\gamma^{5})\nu_{e} + W_{\mu}^{+}\overline{\nu_{e}}\gamma^{\mu}(1-\gamma^{5})e]
\\
&\phantom{{}={}} - \frac{g}{2C_{W}}Z_{\mu}[\frac{1}{2}\overline{\nu_{e}}\gamma^{\mu}(1-\gamma^{5})\nu_{e} + \overline{e}\gamma^{\mu}(-\frac{1}{2}+2S_{W}^{2}+\frac{1}{2}\gamma^{5})e]
\end{split}
\label{eqn:10}
\end{equation}
 
\begin{equation}
L_{int}^{isosinglet}=eA_{\mu}\overline{E}\gamma^{\mu}E-\frac{2gS_{W}^{2}}{2C_{W}}Z_{\mu}\overline{E}\gamma^{\mu}E
\label{eqn:11}
\end{equation}
 
\begin{equation}
\begin{split}
L_{int}^{isodoublet} 
& = eA_{\mu}\overline{E}\gamma^{\mu}E - \frac{g}{\sqrt{2}}[W_{\mu}^{-}\overline{E}\gamma^{\mu}N + W_{\mu}^{+}\overline{N}\gamma^{\mu}E]
\\
&\phantom{{}={}} - \frac{g}{2C_{W}}Z_{\mu}[\overline{N}\gamma^{\mu}N + \overline{E}\gamma^{\mu}(-1+2S_{W}^{2})E]
\end{split}
\label{eqn:12}
\end{equation}
In Equations (12)-(14), mixings of SM and new leptons are neglected. Actually, leptons in these equations should be denoted as $e^0$, $\nu^0$, $N^0$ and $E^0$ which are related to mass eigenstate according to:

\begin{equation}
\begin{aligned}
  e_L^0 &= c_L^E e_L + s_L^E E_L,   &    e_R^0 &= c_R^E e_R + s_R^E E_R\\
  E_L^0 &= -s_L^E e_L + c_L^E E_L,   &    E_R^0 &= -s_R^E e_R + c_R^E E_R\\
\nu_L^0 &= c_L^N \nu_L + s_L^N N_L,   &    \nu_R^0 &= c_R^N \nu_R + s_R^N N_R\\
  N_L^0 &= -s_L^N \nu_L + c_L^N N_L,   &    N_R^0 &= -s_R^N \nu_R + c_R^N N_R
\end{aligned}
\label{eqn:13}
\end{equation}
Lagrangians with mixings are given in \autoref{sec:appendix}.\\\\
In order to calculate production cross-sections and decay rates, we implemented Lagrangians
with mixings into CompHEP software \cite{Boos:2004co,pukhov2000comphep}
	\section{Production of vector-like leptons}
In this section we considered pair and single production of new leptons at energy frontier colliders. Numerical calculations are performed for $s_L^E = s_L^N = s_R^E = s_R^N = 0.01$. This assumption is not crucial for pair production, whereas single production cross-sections are proportional to $s^2$. A possible scenario for small mixing angles is discovery of vector-like leptons in pair production processes at $pp$, $e^+e^−$ and $\mu^+\mu^−$ colliders. Then, a search for single production of discovered lepton at different colliders will give opportunity to determine values of mixing angles or put upper limit on them.
\begin{table}[H]
\centering
\caption{Collider Parameters}
\begin{tabular}{|cc|c|c|}
\hline
\multicolumn{2}{|c|}{\textbf{Collider}}                          & $\mathbf{\sqrt{s} [TeV]}$ & $\mathbf{\mathcal{L}^{int} [ab^{-1}]}$ \\ \hline
\multicolumn{2}{|c|}{HL-LHC \cite{Apollinari:2120673}}                                     & 14          & 3             \\ \hline
\multicolumn{2}{|c|}{HE-LHC \cite{Zimmermann:2651305}}                                     & 27          & 10            \\ \hline
\multicolumn{2}{|c|}{FCC \cite{1c765d761de4411ea2b0f3793185c1c3}}                                        & 100         & 20            \\ \hline
\multicolumn{2}{|c|}{ILC \cite{behnke2013international}}                                        & 1           & 4.9           \\ \hline
\multicolumn{2}{|c|}{CLIC \cite{Burrows:2652188}}                                       & 3           & 5             \\ \hline
\multicolumn{2}{|c|}{PWFA-LC \cite{adli2013beam}}                                    & 10          & 10            \\ \hline
\multicolumn{2}{|c|}{\multirow{4}{*}{MC \cite{delahaye2019muon,Zimmermann:2018wfu}}}                        & 1.5         & 1.25          \\ \cline{3-4} 
\multicolumn{2}{|c|}{}                                           & 3           & 4.4           \\ \cline{3-4} 
\multicolumn{2}{|c|}{}                                           & 6           & 12            \\ \cline{3-4} 
\multicolumn{2}{|c|}{}                                           & 14          & 10            \\ \hline
\multicolumn{1}{|c|}{\multirow{2}{*}{HL-LHC based ep}} & ILC     & 3.7         & 0.02          \\ \cline{2-4} 
\multicolumn{1}{|c|}{}                                 & PWFA-LC & 11.8        & 0.04          \\ \hline
\multicolumn{1}{|c|}{\multirow{2}{*}{HE-LHC based ep}} & ILC     & 5.2         & 0.06          \\ \cline{2-4} 
\multicolumn{1}{|c|}{}                                 & PWFA-LC & 16.4        & 0.12          \\ \hline
\multicolumn{1}{|c|}{\multirow{2}{*}{FCC based ep \cite{Acar:2017ni}}}    & ILC     & 10          & 0.15          \\ \cline{2-4} 
\multicolumn{1}{|c|}{}                                 & PWFA-LC & 31.6        & 0.43          \\ \hline
\end{tabular}
\label{tbl:2}
\end{table}
\noindent
Center of mass energies and integrated luminosities of relevant colliders are presented in Table 2. It should be mentioned that linear colliders (ILC, CLIC and PFWA-LC) allow construction of $\gamma e$ and $\gamma\gamma$ colliders with slightly lower center-of mass energies and approximately similar luminosities (see \cite{Telnov:2016pp} and references therein). Vector-like leptons will be pairly produced at pp, lepton and $\gamma\gamma$ colliders if their masses lie in kinematically allowed region. Single production of VLLs can be observed at $pp$, $ep$, lepton and $\gamma e$ colliders.
        \subsection{Pair Production}
In this subsection we present cross-sections for pair-production of vector-like leptons at proton, electron-positron, photon and muon colliders.
            \subsubsection{Proton colliders}
At proton colliders, both iso-singlet and iso-doublet charged leptons are produced pairly via following Feynman diagrams:
\begin{figure}[H]
  \centering
    \includegraphics[width=0.60\textwidth]{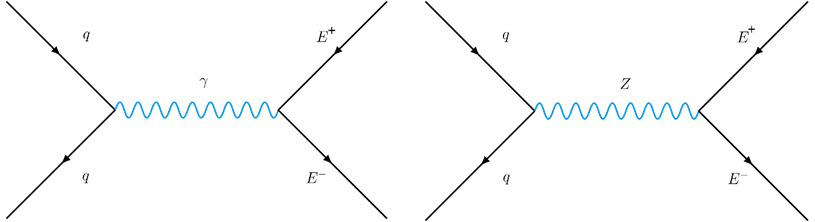}
\caption{Feynman diagrams for pair production of charged VLLs at the LHC/FCC.}
\end{figure}
\noindent
As for neutral vector-like leptons, pair production in iso-singlet case is strongly suppressed by factor $(s_{L,R}^N)^4$. Feynman diagram for iso-doublet case is given below:
\begin{figure}[H]
  \centering
    \includegraphics[width=0.30\textwidth]{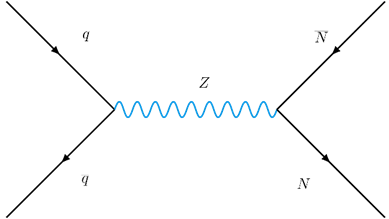}
\caption{Feynman diagram for pair production of neutral VLLs at the LHC/FCC.}
\end{figure}
\noindent
In iso-doublet case, one also deals with associate production of charged and neutral vector-like leptons via diagrams:
\begin{figure}[H]
  \centering
    \includegraphics[width=0.60\textwidth]{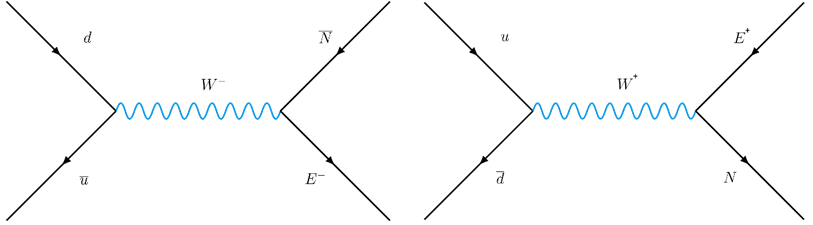}
\caption{Feynman diagrams for associate production of charged and neutral VLLs at the LHC/FCC.\\}
\end{figure}
\noindent
Cross-sections for pair productions of vector-like leptons at the LHC, HE-LHC and FCC are shown in Figure 4 (iso-singlet) and Figures 5-8 (iso-doublet), respectively.
\begin{figure}[H]
  \centering
    \includegraphics[width=0.60\textwidth]{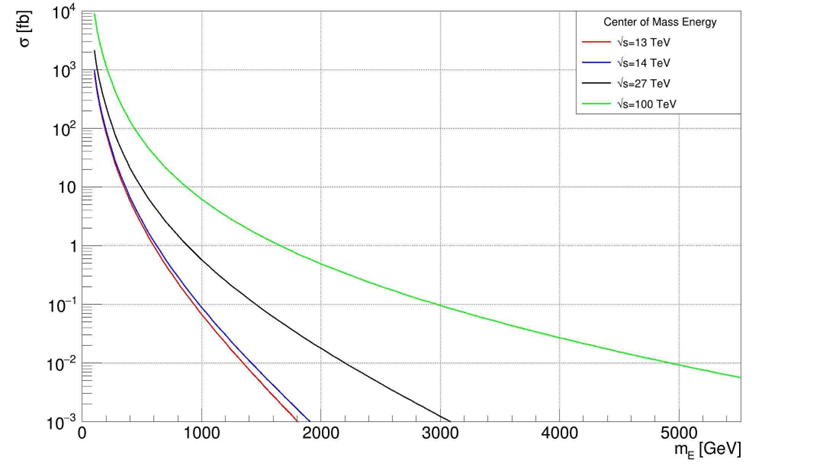}
\caption{Cross-sections for pair production of iso-singlet charged VLLs at the LHC/FCC.}
\end{figure}
\begin{figure}[H]
  \centering
    \includegraphics[width=0.60\textwidth]{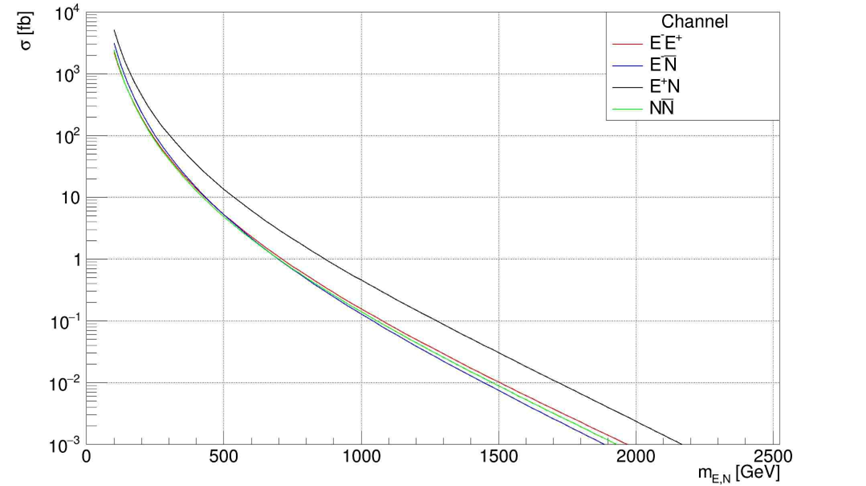}
\caption{Cross-sections for pair production of VLLs at the LHC with $\sqrt{s} = 13$ TeV.}
\end{figure}
\begin{figure}[H]
  \centering
    \includegraphics[width=0.60\textwidth]{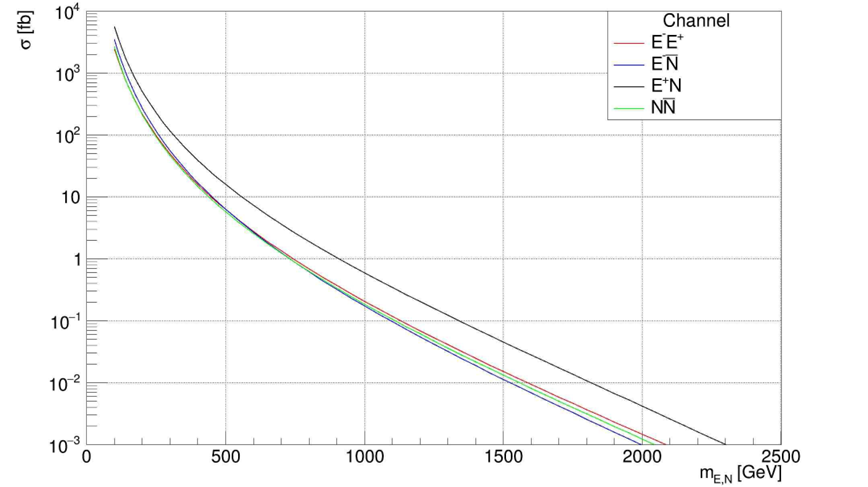}
\caption{Cross-sections for pair production of VLLs at the LHC with $\sqrt{s} = 14$ TeV.}
\end{figure}
\begin{figure}[H]
  \centering
    \includegraphics[width=0.60\textwidth]{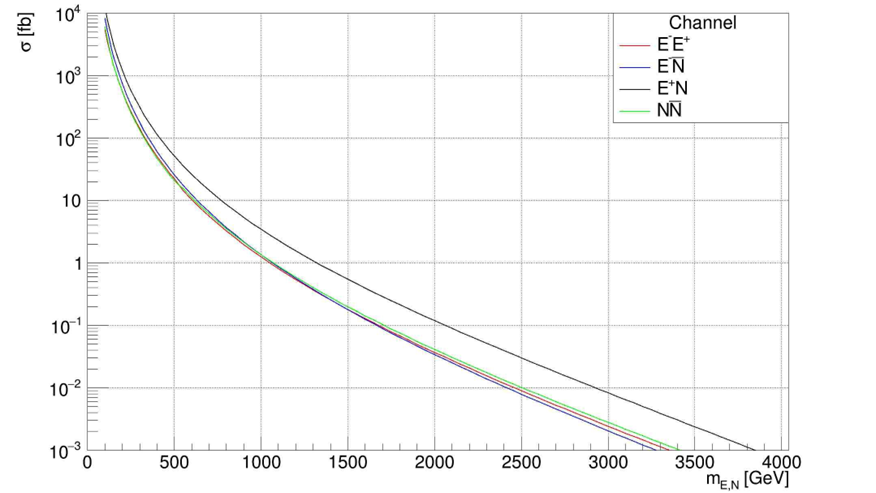}
\caption{Cross-sections for pair production of VLLs at the HE-LHC with $\sqrt{s} = 27$ TeV.}
\end{figure}
\begin{figure}[H]
  \centering
    \includegraphics[width=0.60\textwidth]{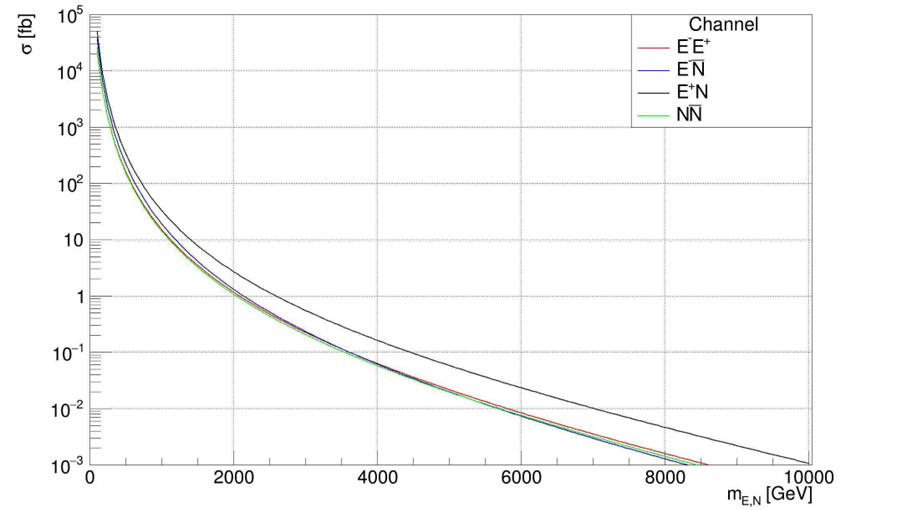}
\caption{Cross-sections for pair production of VLLs at the FCC with $\sqrt{s} = 100$ TeV.}
\end{figure}

            \subsubsection{Electron-positron colliders}
Feynman diagrams for pair production of vector-like leptons at $e^+e^-$ colliders are presented below:
\begin{figure}[H]
  \centering
    \includegraphics[width=0.90\textwidth]{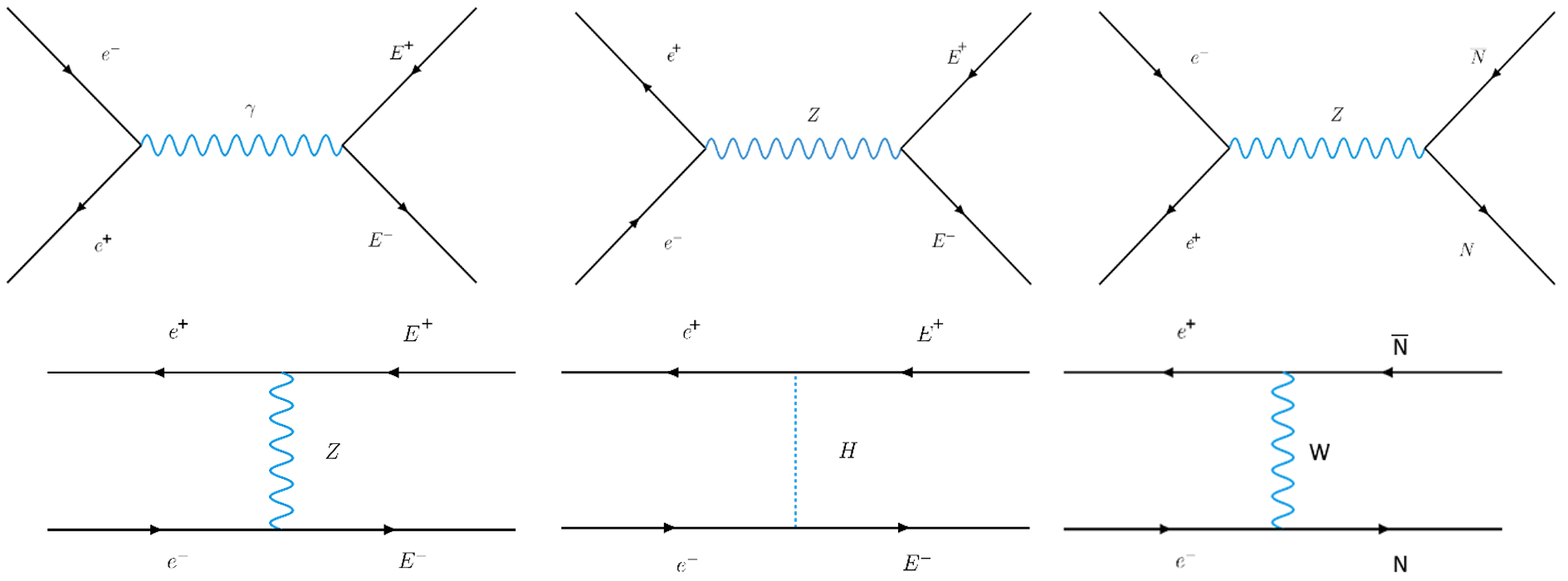}
\caption{Feynman diagrams for pair production of VLLs at $e^+e^-$ colliders.}
\end{figure}

\noindent
 Let us note that pair production of iso-singlet neutral vector-like lepton is suppressed by factor $(s_{L}^N)^4$. Cross-sections for pair production of vector-like leptons at ILC, CLIC and PWFA-LC are presented in Figures 10-12.
\begin{figure}[H]
  \centering
    \includegraphics[width=0.60\textwidth]{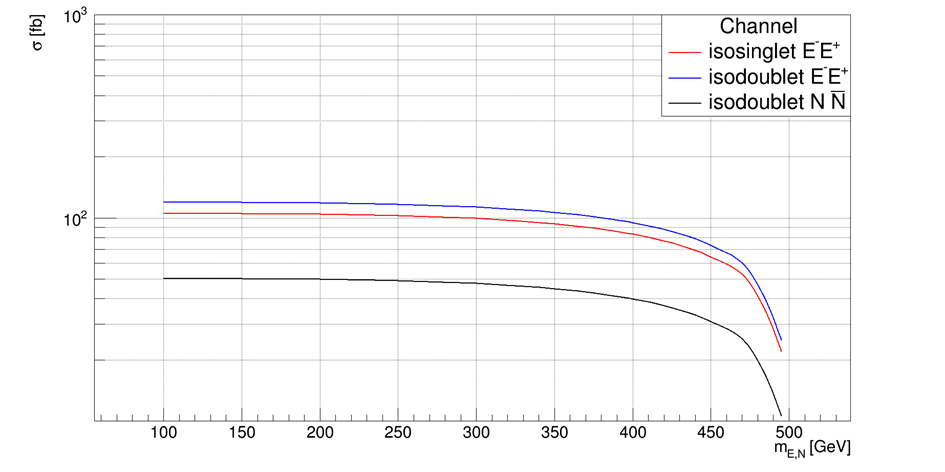}
\caption{Cross-sections for pair production of VLLs at the ILC with $\sqrt{s} = 1$ TeV.}
\end{figure}
\begin{figure}[H]
  \centering
    \includegraphics[width=0.60\textwidth]{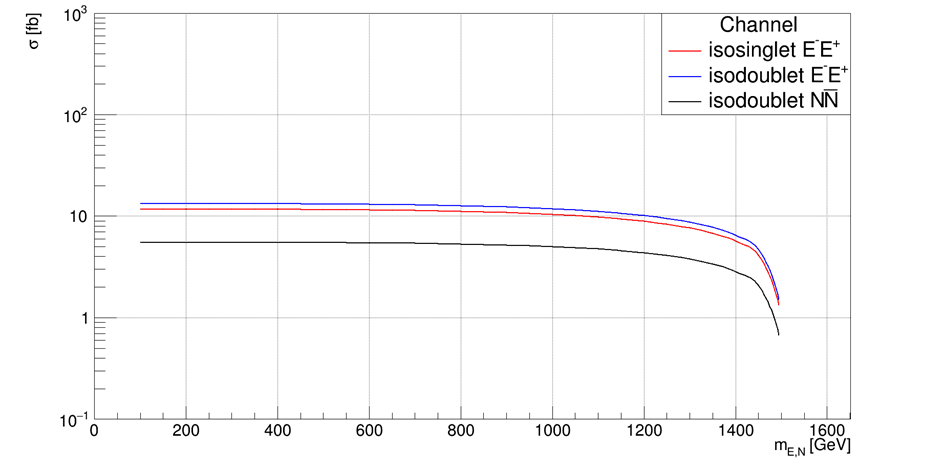}
\caption{Cross-sections for pair production of VLLs at the CLIC with $\sqrt{s} = 3$ TeV.}
\end{figure}
\begin{figure}[H]
  \centering
    \includegraphics[width=0.60\textwidth]{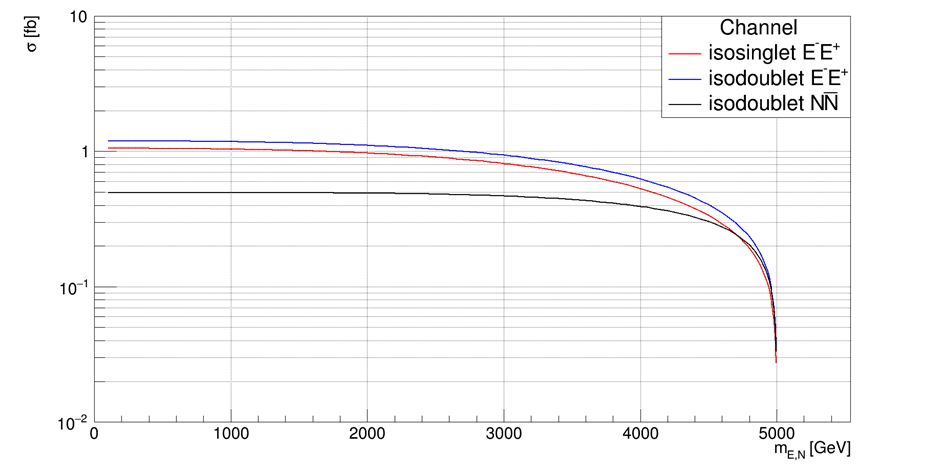}
\caption{Cross-sections for pair production of VLLs at the PWFA-LC with $\sqrt{s} = 10$ TeV.}
\end{figure}
            \subsubsection{Muon colliders}
Feynman diagrams for muon colliders are similar to that of $e^+e^-$ colliders, excepting Z and Higgs bosons exchange diagrams, which are absent since we assume dominance of interfamily mixings with the first SM family leptons. Corresponding cross-sections for muon colliders with $\sqrt{s} = 1.5, 3, 6$ and 14 TeV are pictured in Figures 13-16.
\begin{figure}[H]
  \centering
    \includegraphics[width=0.60\textwidth]{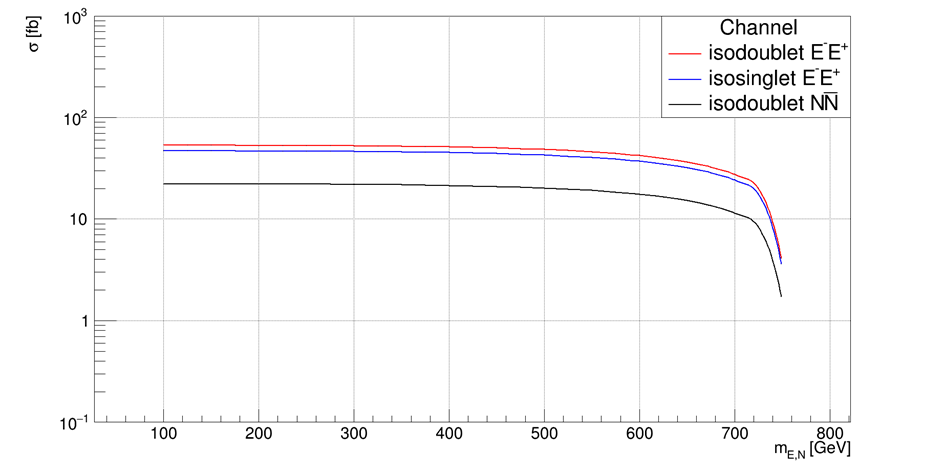}
\caption{Cross-sections for pair production of VLLs at the MC with $\sqrt{s} = 1.5$ TeV.}
\end{figure}
\begin{figure}[H]
  \centering
    \includegraphics[width=0.60\textwidth]{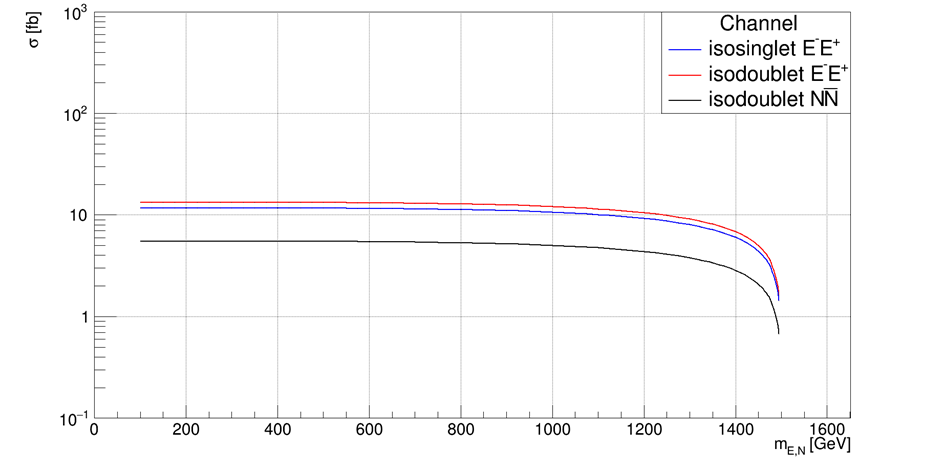}
\caption{Cross-sections for pair production of VLLs at the MC with $\sqrt{s} = 3$ TeV.}
\end{figure}
\begin{figure}[H]
  \centering
    \includegraphics[width=0.60\textwidth]{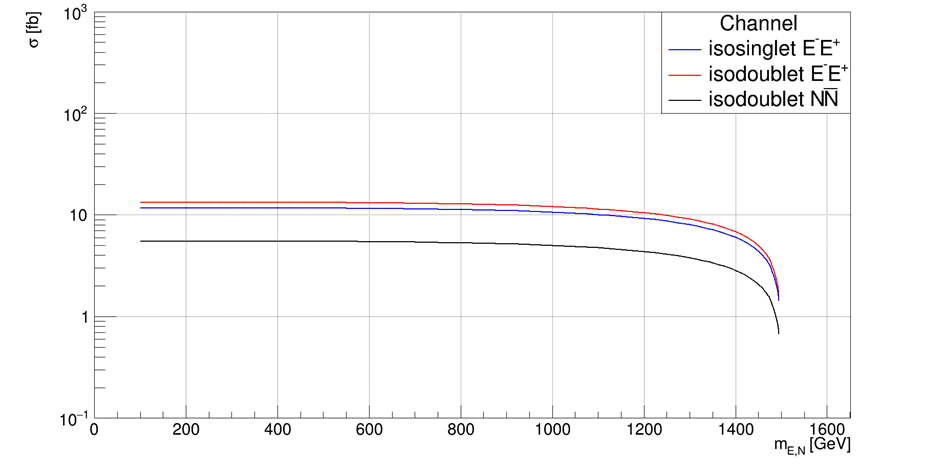}
\caption{Cross-sections for pair production of VLLs at the MC with $\sqrt{s} = 6$ TeV.}
\end{figure}
\begin{figure}[H]
  \centering
    \includegraphics[width=0.60\textwidth]{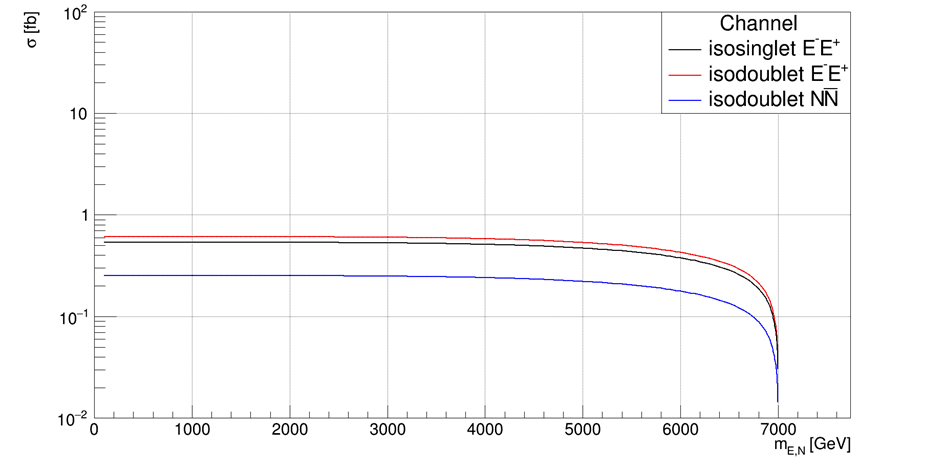}
\caption{Cross-sections for pair production of VLLs at the MC with $\sqrt{s} = 14$ TeV.}
\end{figure}
            \subsubsection{Photon colliders}
Feynman diagrams for pair production of VLLs at $\gamma\gamma$ colliders are shown below:\\
\begin{figure}[H]
  \centering
    \includegraphics[width=0.60\textwidth]{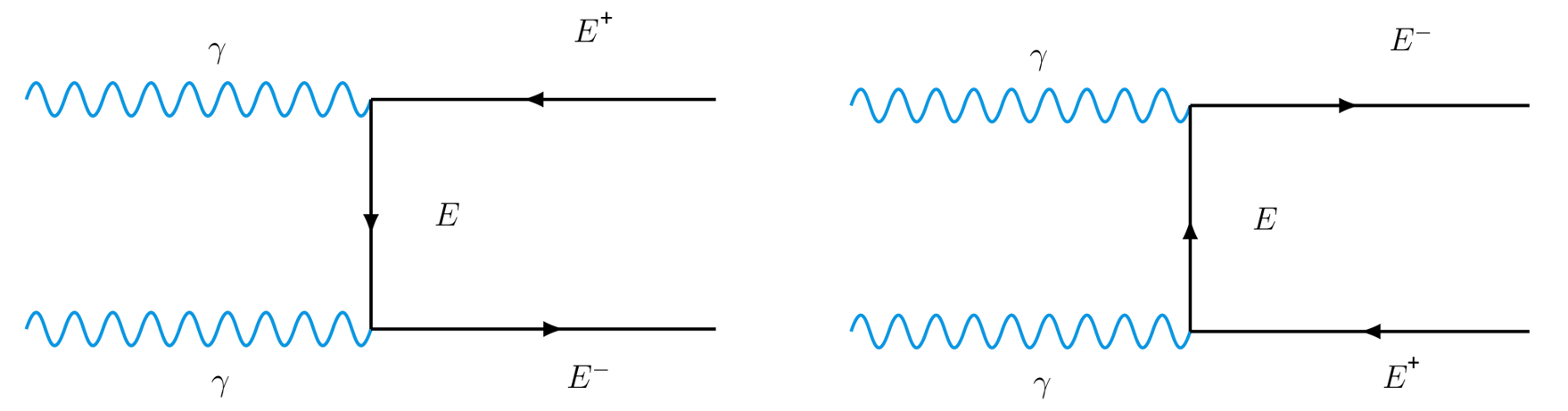}
\caption{Feynman diagrams for pair production of VLLs at photon colliders.}
\end{figure}
\noindent
Corresponding cross-sections at the ILC, CLIC and PFWA-LC based $\gamma\gamma$ colliders are presented in Figures 18-20.
\begin{figure}[H]
  \centering
    \includegraphics[width=0.60\textwidth]{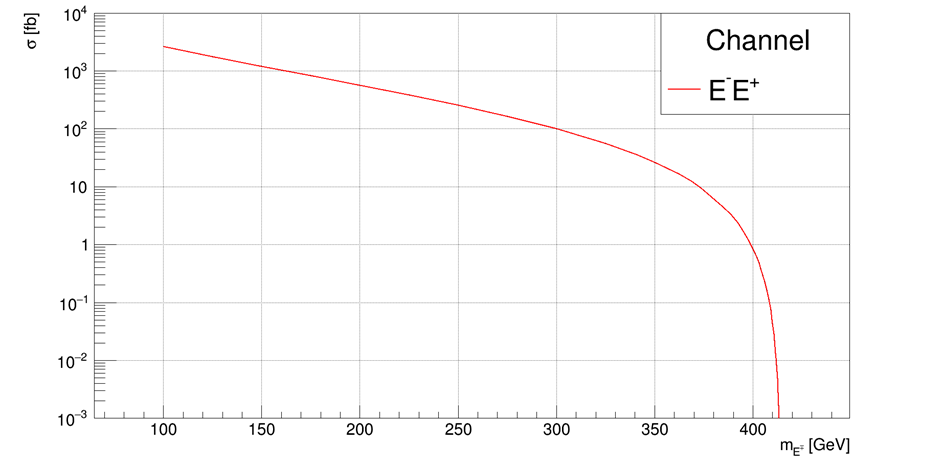}
\caption{Cross-sections for pair production of VLLs at the ILC based $\gamma\gamma$ colliders.}
\end{figure}
\begin{figure}[H]
  \centering
    \includegraphics[width=0.60\textwidth]{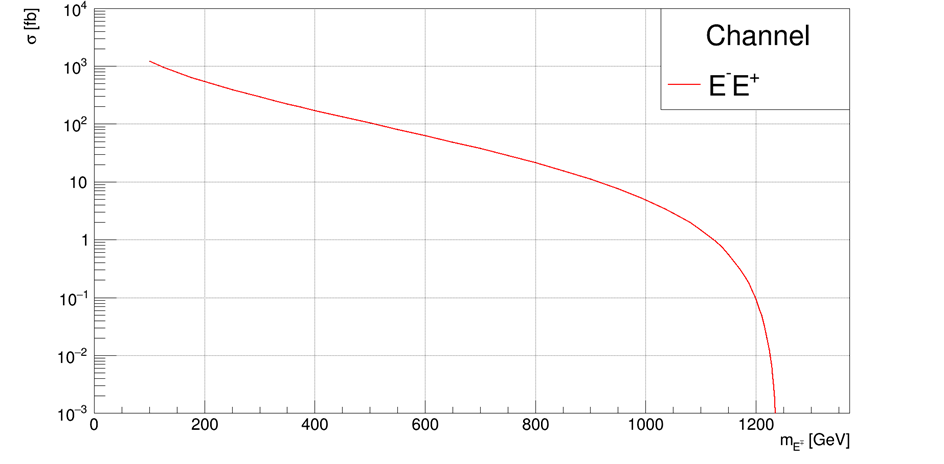}
\caption{Cross-sections for pair production of VLLs at the CLIC based $\gamma\gamma$ colliders.}
\end{figure}
\begin{figure}[H]
  \centering
    \includegraphics[width=0.60\textwidth]{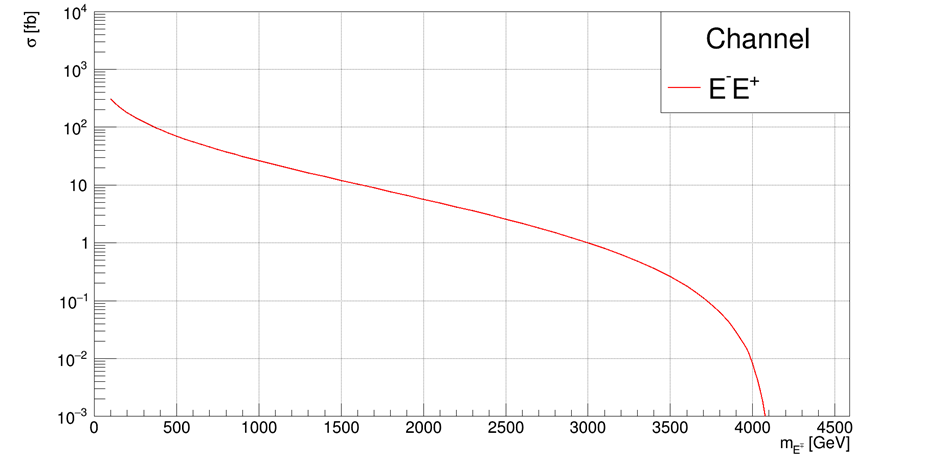}
\caption{Cross-sections for pair production of VLLs at the PWFA-LC based $\gamma\gamma$ colliders.}
\end{figure}

\noindent
Numbers of events for pair production of vector-like leptons with $M_E = M_N = 500$ GeV at different colliders are presented in Table 3. At the ILC (and ILC-$\gamma\gamma$) pair production of VLLs with 0.5 TeV mass is kinematically forbidden. Pair  $\overline{N}N$ production cross-sections in iso-singlet case are negligible since they are proportional to $(s_{L}^{N})^{4}$.
\begin{table}[H]
\centering
\caption{Number of events for VLLs pair production at $m_{E,N} = 500$ GeV. Values without(with) brackets correspond to iso-doublet(iso-singlet) VLLs}
\begin{tabular}{|c|c|c|c|c|}
\hline
\textbf{Collider} & $\mathbf{E^{+}E^{-}}$ & $\mathbf{\overline{N}N}$ & $\mathbf{E^{+}N}$ & $\mathbf{E^{-}\overline{N}}$ \\
\hline
HL-LHC & 21,500 (9,100) & 19,900 (0) & 54,400 (0) & 21,800 (0) \\
\hline
HE-LHC & 231,000(97,300) & 217,000 (0) & 532,000 (0) & 267,000 (0) \\
\hline
FCC & 2,710,000 (1,110,000) & 2,600,000 (0) & 5,710,000 (0) & 3,710,000 (0) \\
\hline
ILC & 0(0) & 0(0) & - & - \\
\hline
CLIC & 66,000 (58,100) & 27,300 (0) & - & - \\
\hline
PWFA-LC & 11,900 (10,500) & 4,900 (0) & - & - \\
\hline
MC (1.5 TeV) & 60,700 (53,400) & 25,300 (0) & - & - \\
\hline
MC (3 TeV) & 58,300 (51,300) & 24,200 (0) & - & - \\
\hline
MC (6 TeV) & 39,900 (35,100) & 16,500 (0) & - & - \\
\hline
MC (14 TeV) & 6,100 (5,400) & 2,500 (0) & - & - \\
\hline
ILC-$\gamma\gamma$ & 0(0) & - & - & - \\
\hline
CLIC-$\gamma\gamma$ & 520,000(520,000) & - & - & - \\
\hline
PWFA-LC-$\gamma\gamma$ & 702,000(702,000) & - & - & - \\
\hline
\end{tabular}
\label{tbl:3}
\end{table}

	    \subsection{Single Production}
In this subsection we consider single production of vector-like leptons at energy frontier colliders. Let's remember that numerical calculations are performed for $s_{L}^{E} = s_{L}^{N} = s_{R}^{E} = s_{R}^{N} = 0.01$. It should be noted that observation of single production of VLLs will give opportunity to determine $s^E$ and $s^N$ values. Moreover, with polarized lepton beams contributions of $s_L$ and $s_R$ may be measured separately.
	        \subsubsection{Electron-proton colliders}
Feynman diagrams for production of vector-like leptons at ep colliders are given below:\\\
\begin{figure}[H]
    \centering
    \includegraphics[width=0.60\textwidth]{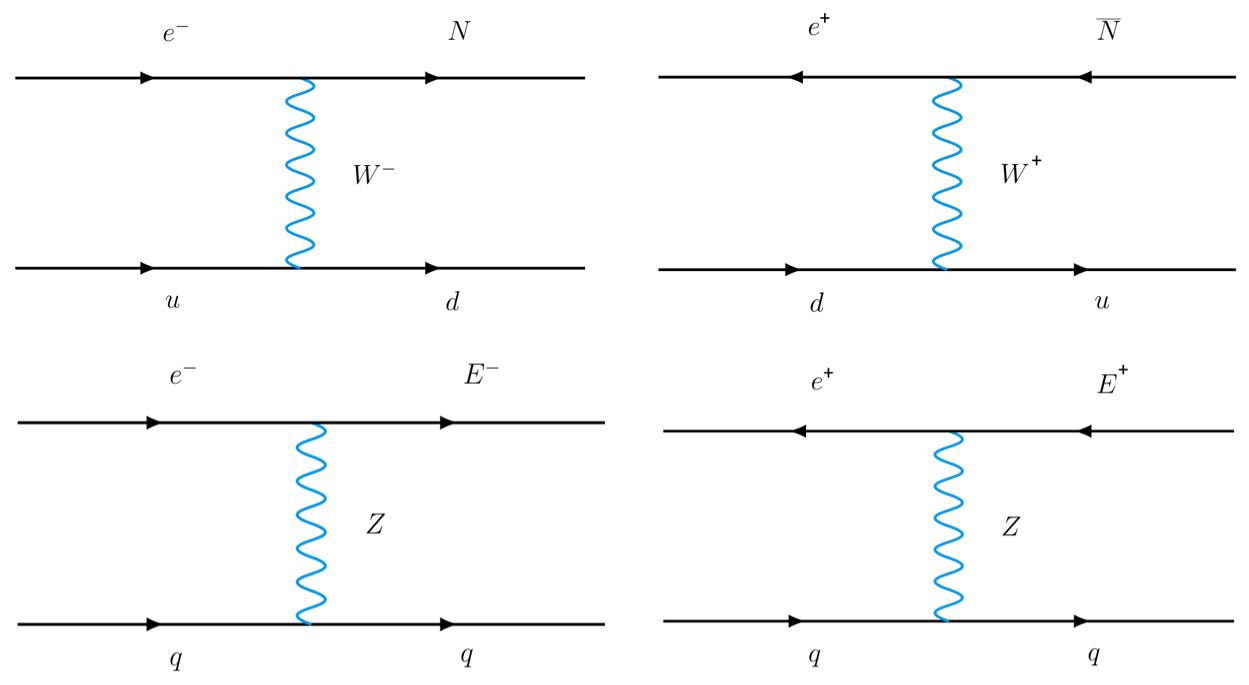}
    \caption{Feynman diagrams for single production of VLLs at ep colliders.}
\end{figure}
\noindent
For illustration we present cross-sections for charged and neutral VLL productions at ILC$\bigotimes$FCC with $\sqrt{s} = 10$TeV.\\\
\begin{figure}[H]
    \centering
    \includegraphics[width=0.60\textwidth]{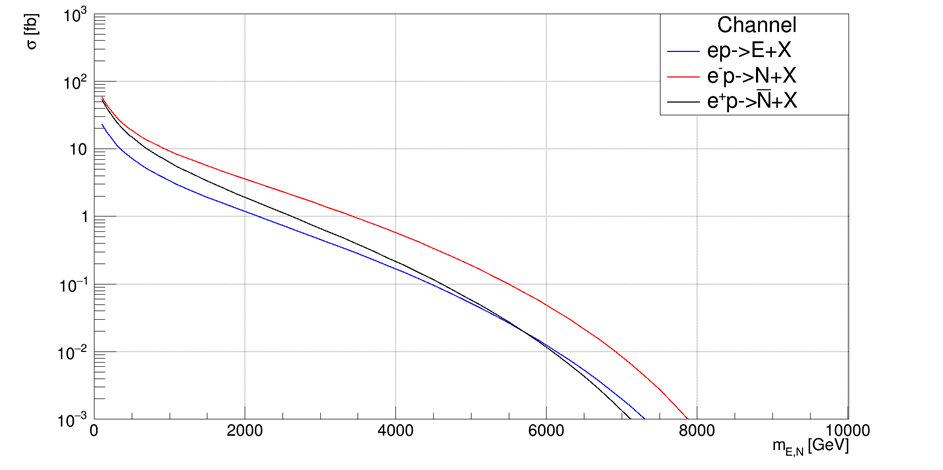}
    \caption{Cross-sections for single production of VLLs at ILC$\bigotimes$FCC with $\sqrt{s} = 10$ TeV.}
\end{figure}
\noindent
Cross-sections for iso-singlet and iso-doublet VLLs are equal since we assume equal values for mixing angles. Let us note that production cross-sections for charged leptons at $e^- p$ and $e^+ p$ colliders are the same. If $M_N = 500$ GeV, production cross-section is $\sigma(e^- p \rightarrow N + X) = 18$ fb, which corresponds to 2700 events for $L^{int} = 0.15$ $ab^{-1}$ (see \autoref{tbl:2}). Remind that in iso-singlet case cross-section is proportional to $(s_L^N)^{2}$ and we used $s_L^N = 0.01$ in numerical calculations. Therefore, in this case one can measure $s_L^N$ down to $10^{-3}$ assuming 25 events as the observation limit. Similar statement is valid for iso-doublet case, where $s_L^N$ is changed to $s_R^E$.
            \subsubsection{Muon-proton colliders}
In the case under consideration (dominant mixing of new leptons with the first SM family leptons) there is no single production at $\mu p$ colliders. If vector-like leptons are predominantly mixed with the second SM family leptons, one deals with situation similar to previous subsection ($\mu$ will replace e, M will replace E).
            \subsubsection{Electron-positron colliders}
Feynman diagrams for single production of vector-like leptons at electron-positron colliders are shown below:\\
\begin{figure}[H]
    \centering
    \includegraphics[width=0.60\textwidth]{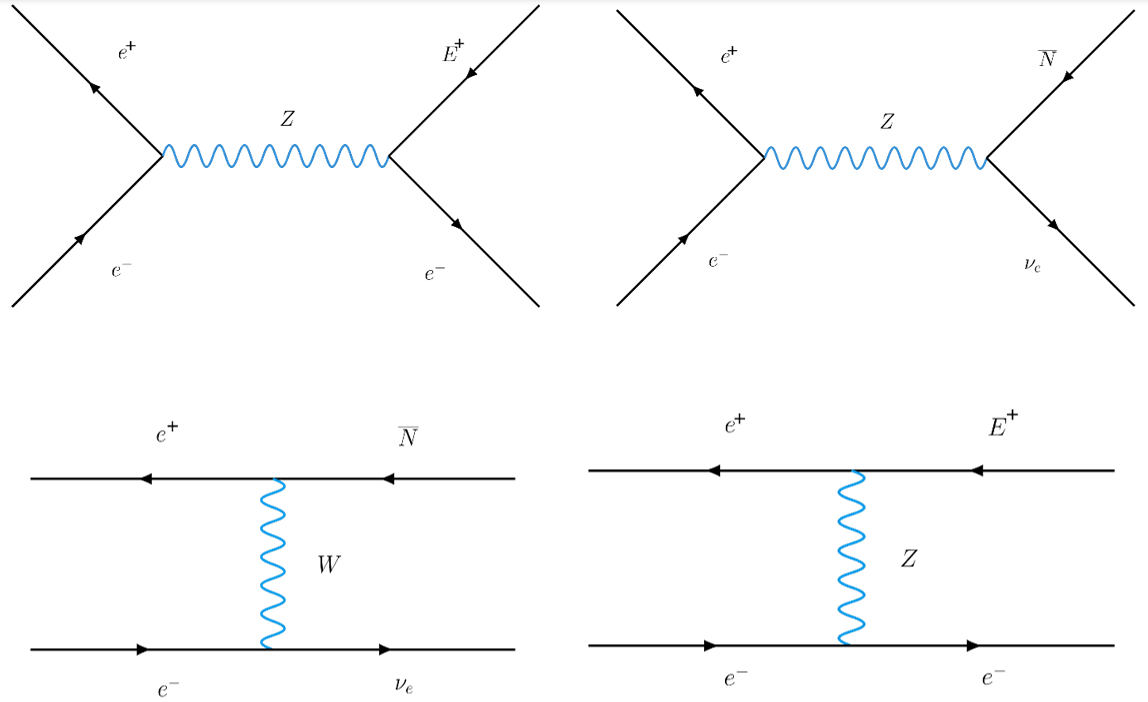}
    \caption{Feynman diagrams for single production of $E^+$ and $\overline{N}$ at $e^+ e^-$ colliders.}
\end{figure}
\noindent
For illustration we present cross-sections for charged and neutral VLL productions at ILC with $\sqrt{s} = 1$ TeV.\\
\begin{figure}[H]
    \centering
    \includegraphics[width=0.60\textwidth]{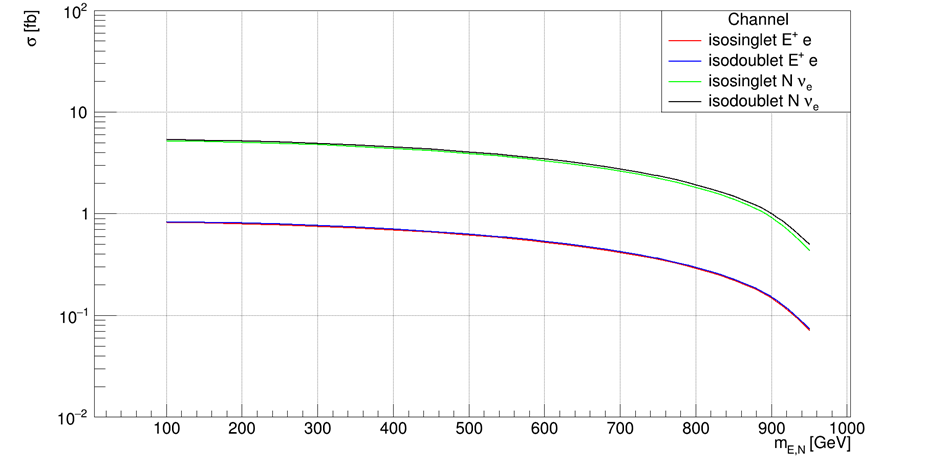}
    \caption{Cross-sections for single production of VLLs at ILC with $\sqrt{s} = 1$ TeV.}
\end{figure}
\noindent
Cross-sections for iso-singlet and iso-doublet VLLs are equal since we assume equal values for mixing angles. Let us note that production cross-sections for $E^+ e^-$ and $E^- e^+$ final states are equal (the same is true for neutral VLLs). If $M_N = 500$ GeV, production cross-section is $\sigma(e^+ e^- \rightarrow \overline{N} \nu_e) = 3.9$ fb, which corresponds to 38200 N and $\overline{N}$ events for $L^{int} = 4.9$ $ab^{-1}$ (see \autoref{tbl:2}). Remind that in iso-singlet case cross-section is proportional to $(s_L^N)^{2}$ (in iso-doublet case cross-section is proportional to $(s_R^E)^{2}$, since diagram with W exchange is dominant) and we used $s_L^N = s_R^E = 0.01$ in numerical calculations. Therefore, in this case one can measure $s_L^N (s_R^E)$ down to $2.6 \times 10^{-4}$ assuming 25 events as the observation limit.
            \subsubsection{Muon colliders}
In the case under consideration there is no single production at muon colliders. If vector-like leptons are predominantly mixed with the second SM family leptons, one deals with situation similar to previous subsection.
            \subsubsection{Proton colliders}
Feynman diagrams for single production of vector-like leptons at proton colliders are given below:\\\
\begin{figure}[H]
    \centering
    \includegraphics[width=0.90\textwidth]{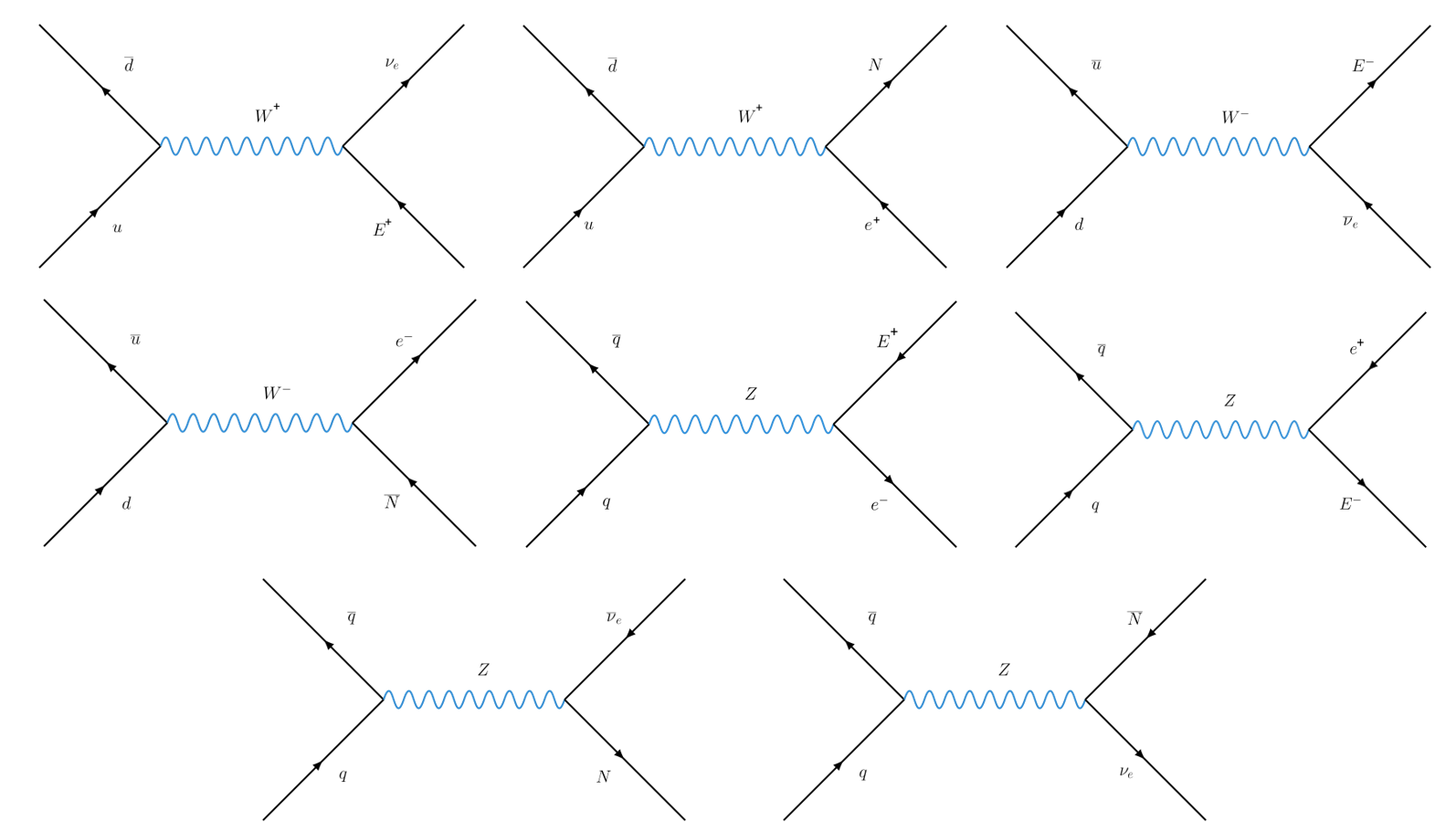}
    \caption{Feynman diagrams for single production of VLLs at proton colliders.}
\end{figure}
\noindent
Cross-sections for single production of VLLs at the LHC and FCC are shown in Figures 26 and 27, respectively. Cross-sections for iso-singlet and iso-doublet VLLs are equal since we assume equal values for mixing angles. Let us remind that production cross-sections are suppressed by factors $(s_L^E )^2$, $(s_L^N )^2$, $(s_R^E )^2$ and $(s_R^N )^2$. Therefore, pair production is much more advantageous in proton collisions. For example, at the HL-LHC $\sigma(pp \rightarrow E^+e^-) = 1.46\times10^{-3} fb$ which corresponds to 4 events at $L^{int} = 3 ab^{-1}$ comparing to 21500 (9100) events for $E^+E^-$ pair production in iso-doublet (iso-singlet) case. Certainly, pp colliders do not have the potential to determine mixing angles, unlike $ep$ and $e^+e^-$ colliders.
\begin{figure}[H]
    \centering
    \includegraphics[width=0.60\textwidth]{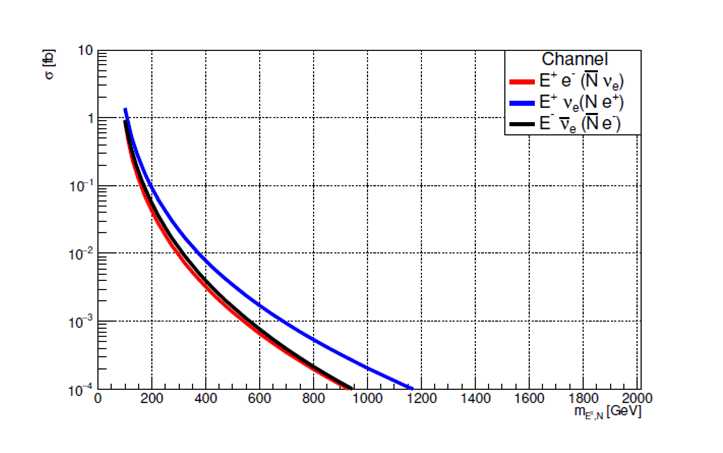}
    \caption{Cross-sections for single production of VLLs at the LHC.}
\end{figure}
\begin{figure}[H]
    \centering
    \includegraphics[width=0.60\textwidth]{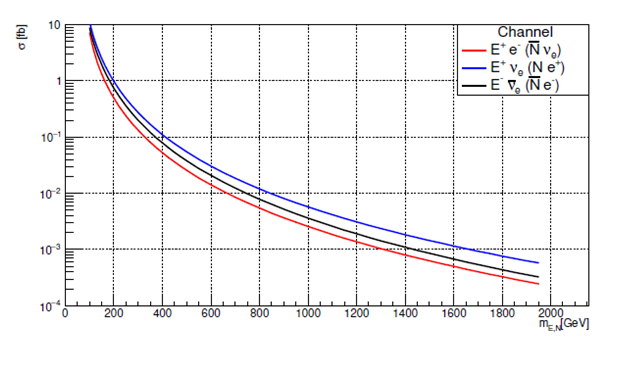}
    \caption{Cross-sections for single production of VLLs at the FCC.}
\end{figure}
            \subsubsection{Photon-electron colliders}
Feynman diagrams for single production of VLLs at ${\gamma}e$ colliders are:\\\ 
\begin{figure}[H]
    \centering
    \includegraphics[width=0.60\textwidth]{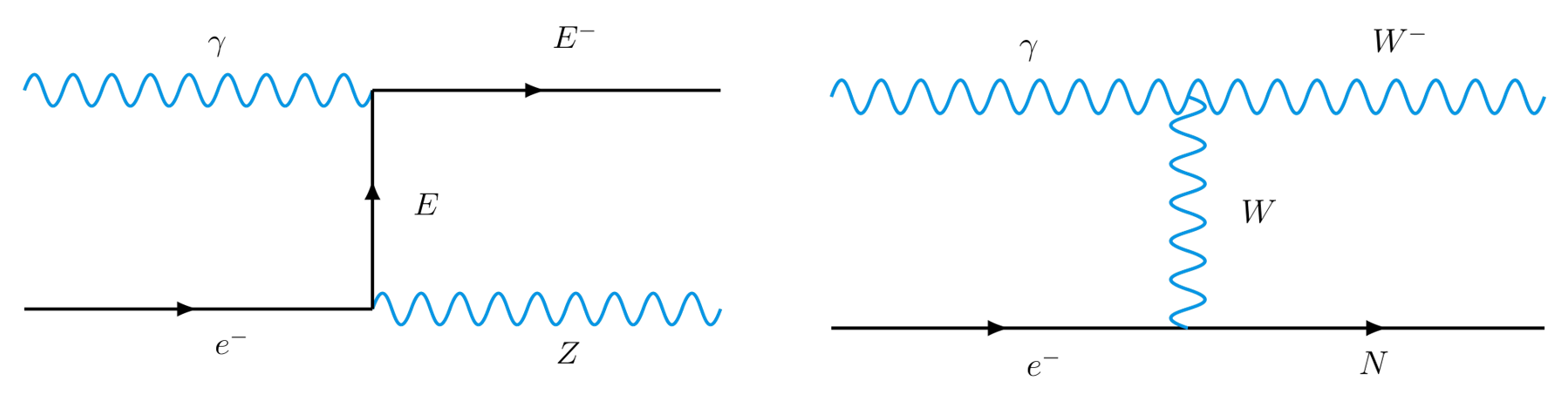}
    \caption{Feynman diagrams for single production of VLLs at photon-electron colliders.}
\end{figure}
\noindent
Corresponding cross-sections at ILC, CLIC and PWFA-LC are presented in Figures 29-31. For instance, let's consider $W^-N$ production at the ILC in details. If $M_N = 500$, GeV production cross-section is $\sigma({\gamma}e^- \rightarrow W^-N) = 1.9$ fb, which corresponds to 9300 events for $L^{int} = 4.9$ $ab^{-1}$. Let us remind that in iso-singlet case cross-section is proportional to $(s_L^N)^2$ and we used $s_L^N=0.01$ in numerical calculations. Therefore, in this case one can measure $s_L^N$ down to $5\times10^{-4}$ assuming 25 events as the observation limit. Similar statement is valid for iso-doublet case, where $s_L^N$ is changed to $s_R^E$. 
\begin{figure}[H]
    \centering
    \includegraphics[width=0.60\textwidth]{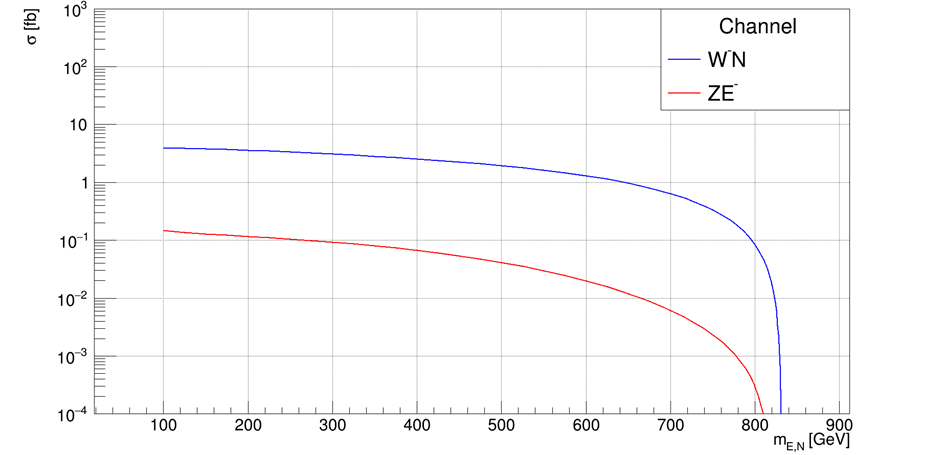}
    \caption{Cross-sections for single production of VLLs at the ILC based $\gamma e$ colliders.}
\end{figure}
\begin{figure}[H]
    \centering
    \includegraphics[width=0.60\textwidth]{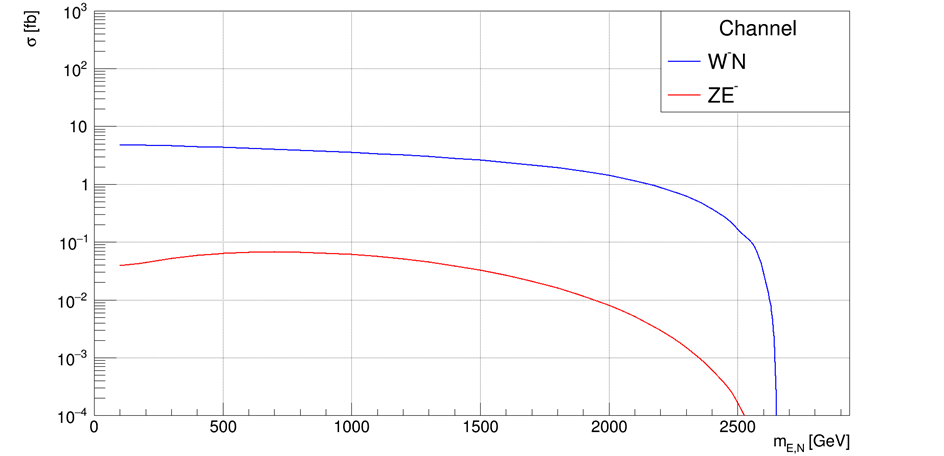}
    \caption{Cross-sections for single production of VLLs at the CLIC based $\gamma e$ colliders.}
\end{figure}
\begin{figure}[H]
    \centering
    \includegraphics[width=0.60\textwidth]{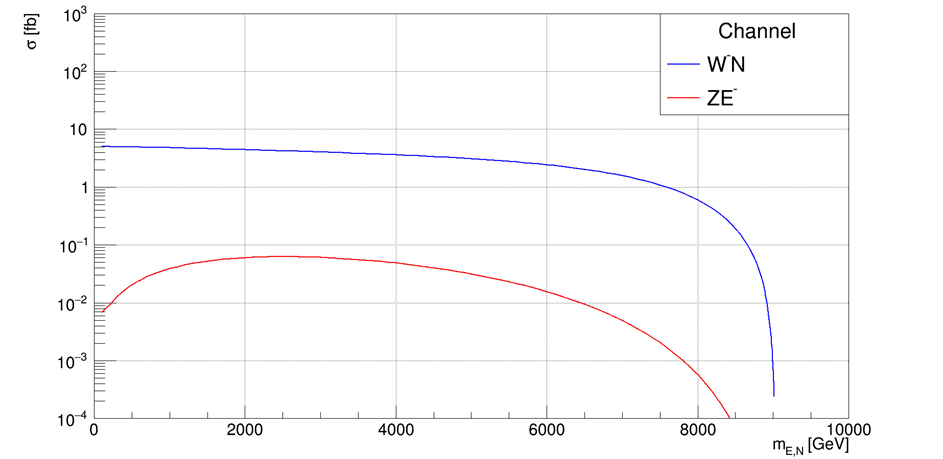}
    \caption{Cross-sections for single production of VLLs at the PWFA-LC based $\gamma e$ colliders.}
\end{figure}
	\section{Vector-like lepton decays}
Decay patterns of vector-like leptons depend on mixing angles and mass relations between charged and neutral vector-like leptons. In this section, we examine possible different decay scenarios. Decay width formulas are calculated using corresponding terms of Lagrangians given in the \autoref{sec:appendix}.

	    \subsection{E is the lightest vector-like lepton}
	    \label{5.1}
If E is the lightest vector-like lepton, it will decay only through mixings with first SM family leptons: $E^{-} \rightarrow W^{-}\nu$, $E^{-} \rightarrow Ze^{-}$ and $E^{-} \rightarrow He^{-}$. In this subsection notation $a_{E} = g^{2}M_{E}^{2}/2m_{W}^{2}$ and $r_{X}^{E} = M_{X}^{2}/M_{E}^{2} \ (X = W, Z, H)$ are used.
	        \subsubsection{Iso-singlet case}
Decay widths calculated for iso-singlet E are given below:	        
\begin{equation}
\Gamma(E^{-} \rightarrow W^{-}\nu) = \frac{M_{E}}{32\pi}a_{E}(c_L^N)^{2}(s_L^E)^{2}(1 - r_W^E)^{2}(1 + 2r_W^E)
\end{equation}
\begin{equation}
\Gamma(E^{-} \rightarrow Ze^{-}) = \frac{M_{E}}{64\pi}a_{E}(c_L^E)^{2}(s_L^E)^{2}(1 - r_Z^E)^{2}(1 + 2r_Z^E)
\end{equation}
\begin{equation}
\Gamma(E^{-} \rightarrow He^{-}) = \frac{M_{E}}{64\pi}a_{E}(c_L^E)^{2}(s_L^E)^{2}(1 - r_H^E)^{2}
\end{equation}
For $M_E >> m_{W,Z}$ branching ratios become $BR(E^{-} \rightarrow W^{-}\nu)  = 0.5$ and $BR(E^{-} \rightarrow Ze^{-}) = BR(E^{-} \rightarrow He^{-}) = 0.25$. Dependence of these branching ratios on the mass of charged vector-like lepton is presented in Figure 32.
\begin{figure}[H]
    \centering
    \includegraphics[width=0.50\textwidth]{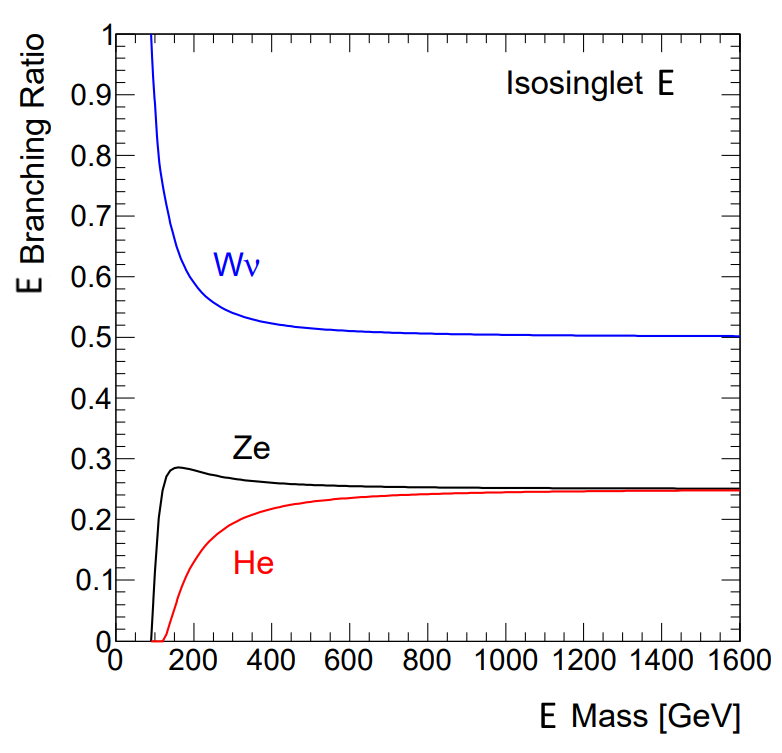}
    \caption{Branching ratios for decays of iso-singlet charged vector-like lepton.}
\end{figure}
\noindent
The decay widths of vector-like leptons are crucial for their experimental observation. In the iso-singlet case, the variation of the decay widths with respect to the charged VLL mass is shown in Figure 33 for the value $s^E_L = 0.01$.
\begin{figure}[H]
    \centering
    \includegraphics[width=0.60\textwidth]{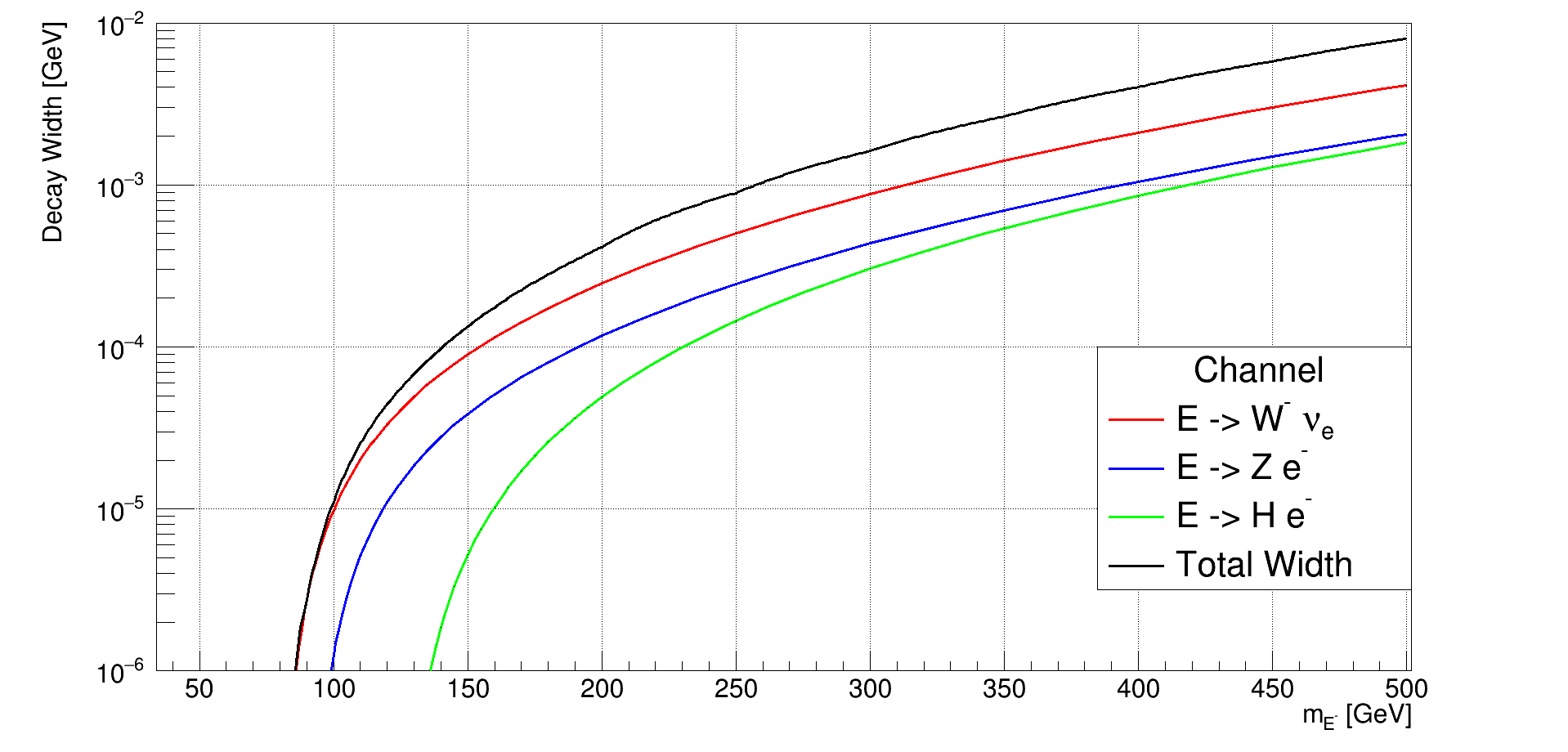}
\end{figure}
\begin{figure}[H]
    \centering
    \includegraphics[width=0.60\textwidth]{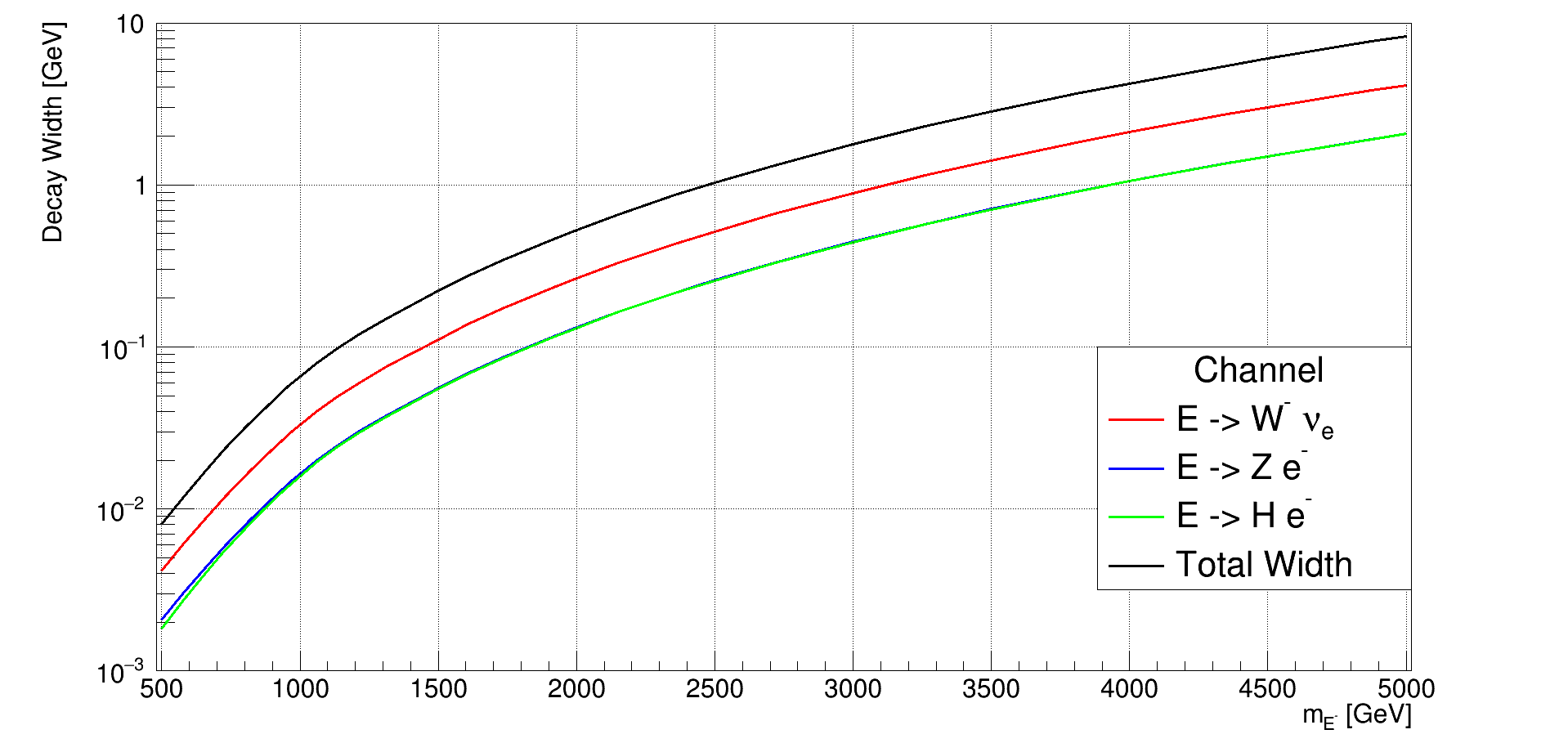}
    \caption{Decay width dependence on the iso-singlet charged VLL mass in the case of the value $s^E_L = 0.01$. }
\end{figure}
\noindent
 It should be noted that the value of decay width is proportional to $(s^E_L)^2$. The decay width become proportional to $M_E^3$ for values exceeding to 1 TeV. It is seen that the decay width value is under experimental resolution even at $M_E =$ 5000 GeV.
	        \subsubsection{Iso-doublet case}
Decay width formulas for iso-doublet E are:
\begin{equation}
\Gamma(E^{-} \rightarrow W^{-}\nu) = \frac{M_{E}}{32\pi}a_{E}[(c_L^N s_L^E - c_L^E s_L^N)^{2}+(c_R^E)^{2}(s_R^N)^{2}](1 - r_W^E)^{2}(1 + 2r_W^E)
\end{equation}
\begin{equation}
\Gamma(E^{-} \rightarrow Ze^{-}) = \frac{M_{E}}{64\pi}a_{E}(c_R^E)^{2}(s_R^E)^{2}(1 - r_Z^E)^{2}(1 + 2r_Z^E)
\end{equation}
\begin{equation}
\Gamma(E^{-} \rightarrow He^{-}) = \frac{M_{E}}{64\pi}a_{E}(c_R^E)^{2}(s_R^E)^{2}(1 - r_H^E)^{2}
\label{eqn:25}
\end{equation}
\noindent
In the $M_E >> m_{W,Z}$ case branching ratios become $BR(E^{-} \rightarrow W^{-}\nu)  = 0.5$ and $BR(E^{-} \rightarrow Ze^{-}) = BR(E^{-} \rightarrow He^{-}) = 0.25$ if $s_L^E = s_L^N$ and $s_R^E = s_R^N$ (Figure 32 is valid for this case as well). 
However, in the case of $s^E_L = s^N_L$ and $s^N_R = 0$, only neutral current decay modes are survived. Dependence of corresponding branching ratios on $M_E$ is presented in Figure 34. Similar situation take place if $s^E_R$ is dominant ($s^E_R >> s^E_L,s^N_L,s^N_R$), since charged current decay mode is suppressed.
\begin{figure}[H]
    \centering
    \includegraphics[width=0.50\textwidth]{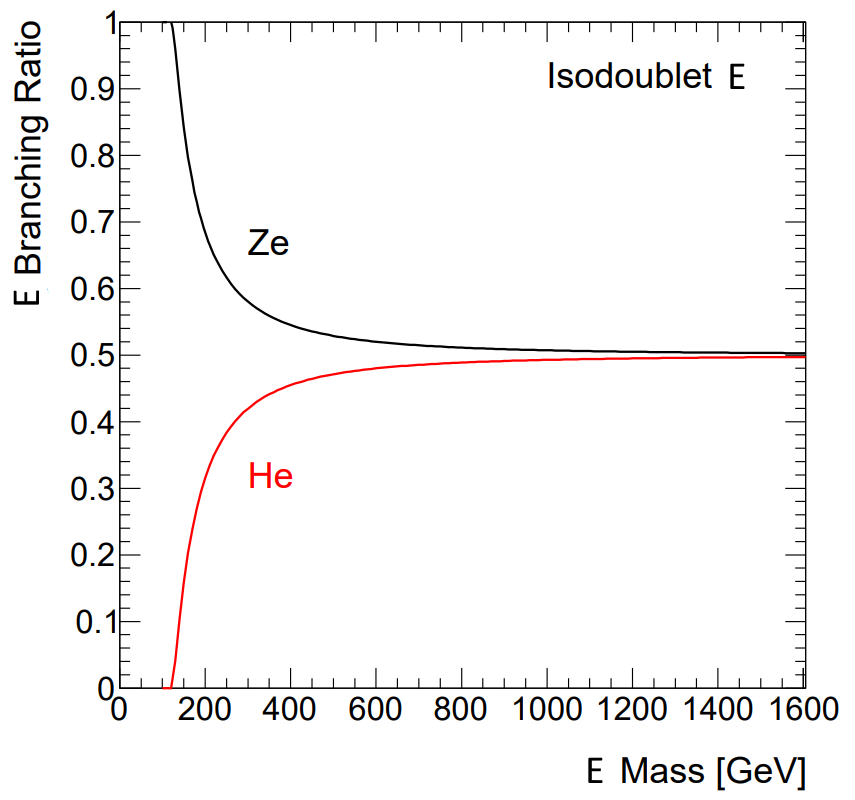}
    \caption{Branching ratios for decays of iso-doublet charged vector-like lepton (neutral currents only).}
\end{figure}
\noindent
The dependence of the decay widths on the charged VLL mass is shown in Figure 35 in the case of only neutral current decay modes survived and $s^E_R = 0.01$. Let us remind that the decay width is proportional to $(s^E_R)^2$ for all masses. The decay width become proportional to $M_E^3$ for values exceeding to 1 TeV. It is seen that the decay width value is under experimental resolution even at $M_E =$ 5000 GeV.
\begin{figure}[H]
    \centering
    \includegraphics[width=0.60\textwidth]{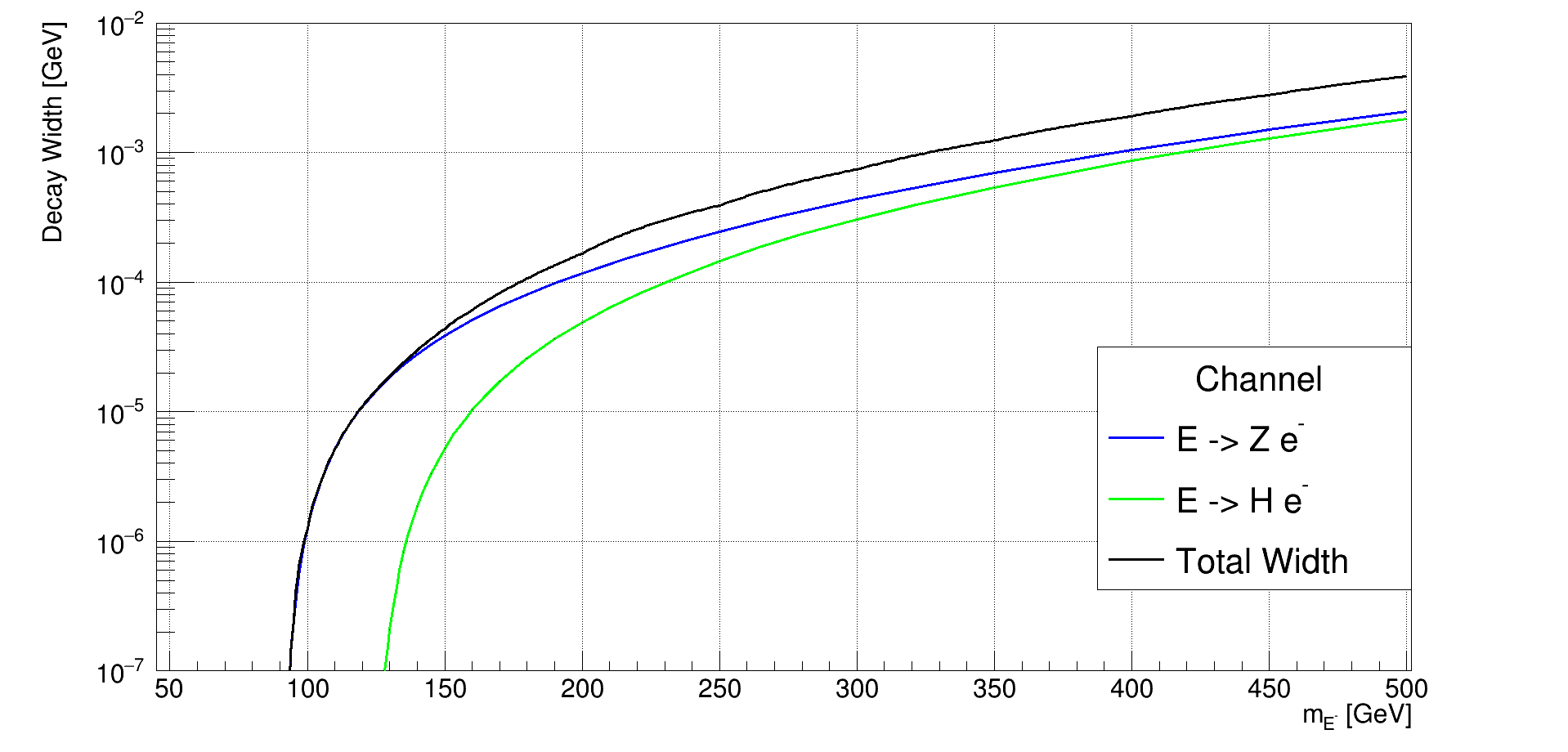}
\end{figure}
\begin{figure}[H]
    \centering
    \includegraphics[width=0.60\textwidth]{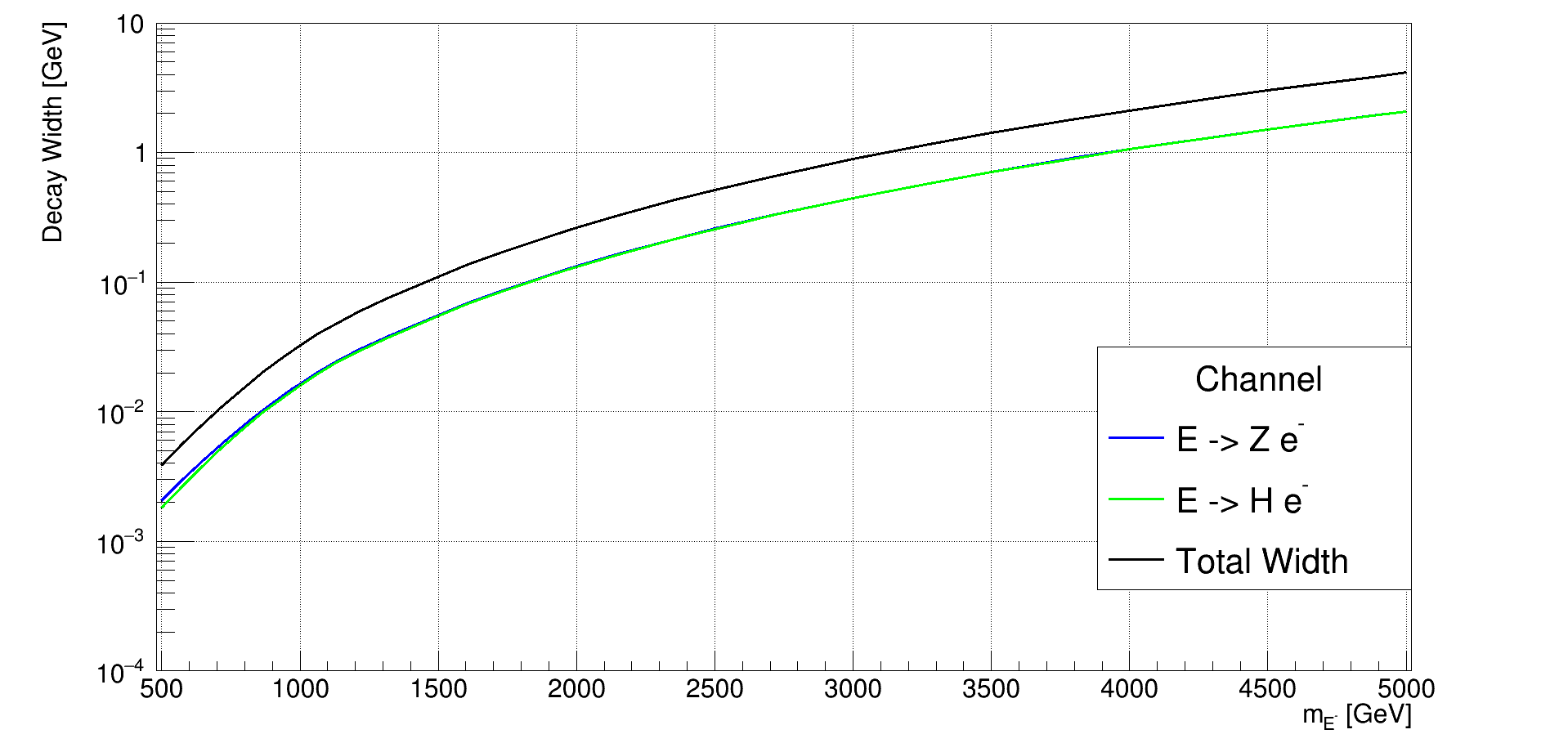}
    \caption{Decay width dependence on the iso-doublet charged VLL mass in the case that $s^E_L=s^N_L$, $s^N_R = 0$ and $s^E_R = 0.01$.}\end{figure}
\noindent
In the case of $s^E_R = 0$ only charged current decay mode survived: $Br(E^{-} \rightarrow W^{-}\nu) = 1$. The mass dependence of the decay width coincides with red line in Figure 33 if one assumes that $s^E_L = s^N_L$ and $s^N_R = 0.01$.

\noindent
In the realistic situation, the mixing angles are expected to be different from each other and from zero. As an example, let us consider the case where right-handed mixings predominate ($s^E_R, s^N_R >> s^E_L, s^N_L$). In this case, charged current decay mode is proportional to $(s^N_R)^2$ and neutral current decay modes are proportional to $(s^E_R)^2$. In Figure 36, we present dependencies of branching ratios on $s^E_R$ for $M_E=1000$ GeV and $s^N_R = 0.01.$
\begin{figure}[H]
    \centering
    \includegraphics[width=0.60\textwidth]{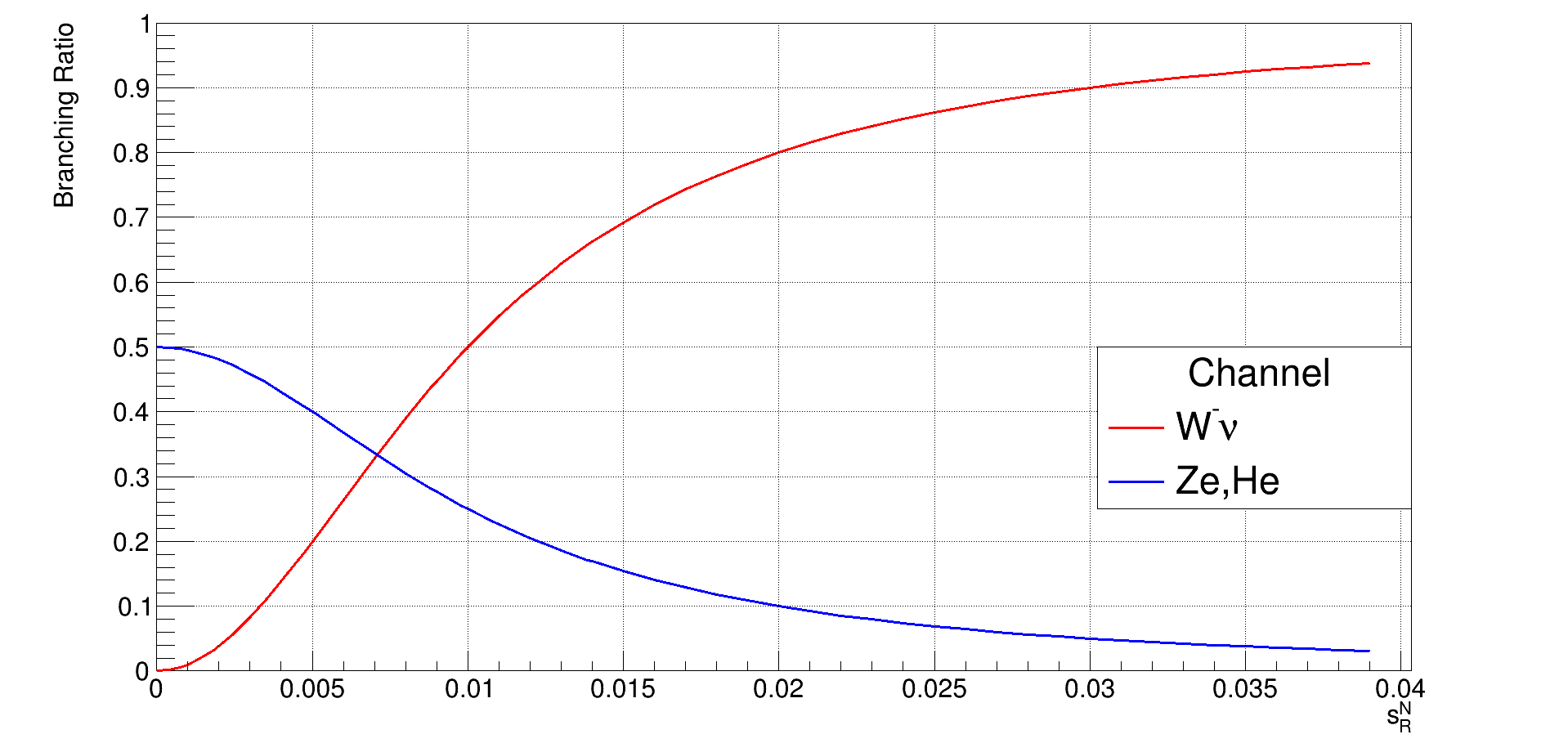}
    \caption{Branching ratios dependencies on $s^E_R$ in the case of $s^N_R = 0.01$, $s^E_R, s^N_R >> s^E_L, s^N_L$ and $M_E=1000$ GeV.}
\end{figure}
\noindent
If left-handed mixings are predominant and $|s^E_L - s^N_L| >> s^{E,N}_R$, the main decay mode is $E^{-} \rightarrow W^{-}\nu$: $Br(E^{-} \rightarrow W^{-}\nu) \approx 1$. In this case decay width is proportional to $(s^E_L - s^N_L)^2$.
        \subsection{N is the lightest vector-like lepton}
        \label{5.2}
If N is the lightest vector-like lepton, it will decay only through mixing with first SM family leptons: $N \rightarrow W^{+}e$, $N \rightarrow Z\nu$ and $N \rightarrow H\nu$. In this subsection notation $a_{N} = g^{2}M_{N}^{2}/2m_{W}^{2}$ and $r_{X}^{N} = M_{X}^{2}/M_{N}^{2} \\ (X = W, Z, H)$ are used.
	        \subsubsection{Iso-singlet case}
Decay widths for this case are given below:
\begin{equation}
\Gamma(N \rightarrow W^{+}e) = \frac{M_{N}}{32\pi}a_{N}(c_L^E)^{2}(s_L^N)^{2}(1 - r_W^N)^{2}(1 + 2r_W^N)
\label{eqn:14}
\end{equation}
\begin{equation}
\Gamma(N \rightarrow Z\nu) = \frac{M_{N}}{64\pi}a_{N}(c_L^N)^{2}(s_L^N)^{2}(1 - r_Z^N)^{2}(1 + 2r_Z^N)
\end{equation}
\begin{equation}
\Gamma(N \rightarrow H\nu) = \frac{M_{N}}{64\pi}a_{N}(c_L^N)^{2}(s_L^N)^{2}(1 - r_H^N)^{2}
\end{equation}
Let us note that, for $M_N >> m_{W,Z}$ branching ratios become $BR(N \rightarrow W^{-}e)  = 0.5$ and $BR(N \rightarrow Z\nu) = BR(N \rightarrow H\nu) = 0.25$. Dependence of these branching ratios on the mass of neutral vector-like lepton is same as charged one which is presented in Figure 32 with replacement of E by N and $e-\nu_{e}$ swapping. Using the same substitutions, other parts of the analysis for the iso-singlet charged VLL performed in section 5.1.1 can also be used for iso-singlet neutral VLL.
	        \subsubsection{Iso-doublet case}
Corresponding formulas for iso-doublet N decays are:
\begin{equation}
\Gamma(N \rightarrow W^{+}e) = \frac{M_{N}}{32\pi}a_{N}[(c_L^N s_L^E - c_L^E s_L^N)^{2}+(c_R^N)^{2}(s_R^E)^{2}](1 - r_W^N)^{2}(1 + 2r_W^N)
\end{equation}
\begin{equation}
\Gamma(N \rightarrow Z\nu) = \frac{M_{N}}{64\pi}a_{N}(c_R^N)^{2}(s_R^N)^{2}(1 - r_Z^N)^{2}(1 + 2r_Z^N)
\end{equation}
\begin{equation}
\Gamma(N \rightarrow H\nu) = \frac{M_{N}}{64\pi}a_{N}(c_R^N)^{2}(s_R^N)^{2}(1 - r_H^N)^{2}
\end{equation}
In the $M_N >> m_{W,Z}$ case branching ratios become $BR(N \rightarrow W^{-}e)  = 0.5$ and $BR(N \rightarrow Z\nu) = BR(N \rightarrow H\nu) = 0.25$ if $s_L^E = s_L^N$ and $s_R^E = s_R^N$. In the case of $s^E_L = s^N_L$ and $s^E_R = 0$, only neutral current decay modes are survived. Dependence of corresponding branching ratios on the mass of neutral vector-like lepton is same as charged one presented in Figure 34 with replacement of E by N and e by $\nu_{e}$. Using the same substitutions, other parts of the analysis for the iso-doublet charged VLL performed in section 5.1.1 can also be used for iso-doublet neutral VLL.

	    \subsection{Iso-doublet: Equal masses}
This case may be considered as natural, if only one vector-like doublet exists. In fact, it is more natural for every family to have their own vector-like partners as in the $E_6$ model. Even in the case of one vector-like doublet, the masses of the charged and uncharged new leptons become different when the SM leptons gain masses (see \autoref{sec:2.1}). If $M_E = M_N$, one can apply formulas given in sections 5.1 and 5.2.\\\\
As mentioned in the introduction, production of vector-like partners of the third SM family leptons at the LHC have been considered in \cite{Kumar:2015vl,Bhattiprolu:2019pf}. Let us compare our results on VLL decays (Equations (16)-(27)) with decay width formulas from \cite{Kumar:2015vl}.\\\\
\noindent
a) Iso-singlet case\\\\
\noindent
Decays of neutral iso-singlet leptons are not included in \cite{Kumar:2015vl}. This corresponds to $s_L^N = 0$ in our equations (22)-(24). Most likely, the authors of \cite{Kumar:2015vl} assumed that neutrinos had no right-handed components (hence, iso-singlet neutral vector-like leptons do not exist). Let us repeat that right-handed neutrinos should be included into SM since there are counterparts of right-handed components of up quarks and observation of neutrino oscillations confirms this statement.\\\\
Concerning charged vector-like leptons, our results (Equations (16)-(18)) overlap with Equations (2.18)-(2.20) in \cite{Kumar:2015vl} substituting e by $\tau$, E by $\tau^{'}$, N by $\nu^{'}$ and taking $\epsilon = \frac{g}{\sqrt{2}}\frac{M_E}{M_W}(s_L^E)$.\\\\
\noindent
b) Iso-doublet case\\\\
\noindent
Our formulas for neutral vector-like lepton decays (Equations (25)-(27)) overlap with Equations (2.28)-(2.29) in \cite{Kumar:2015vl} if $s_R^N = 0$ which result in zero decay width for neutral current decays (let us remind that authors of \cite{Kumar:2015vl} assumed that there are no right-handed components of neutrinos).\\\\
As for charged vector-like lepton decays (Equations (19)-(21)), W channel becomes zero (Equations (2.25) in \cite{Kumar:2015vl}) if $s_L^E = s_L^N = s_R^N = 0$. However, it is unnatural to assume zero value for all three mixings. Decays into Z and H channels overlap with Equations (2.26)-(2.27) in \cite{Kumar:2015vl} if $\epsilon = \frac{g}{\sqrt{2}}\frac{M_E}{M_W}(s_R^E)$.

	    \subsection{Heavier charged lepton}
In this case, additional decay channels are involved, namely, $E^- \rightarrow W^- N$ if $M_E > M_N + m_W$ and $E^- \rightarrow N l^- \Bar{\nu_l}, N q^{'}\Bar{q}$ if $M_N < M_E < M_N + m_W$ (see Figure 37).\\
\begin{figure}[H]
    \centering
    \includegraphics[width=0.40\textwidth]{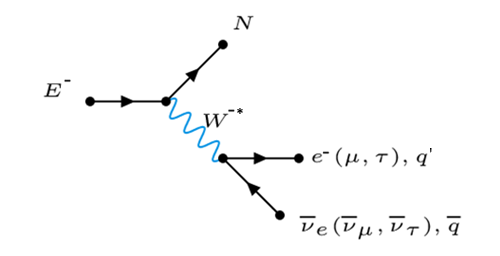}
    \caption{Feynman diagram for heavier charged lepton decay.}
\end{figure}
\noindent
In iso-singlet case, these channels can be neglected since (according to $L^W_{isosinglet}$ in \autoref{sec:appendix}) their decay widths are proportional to $(s_L^N s_L^E)^{2}$. Concerning iso-doublet VLLs unsuppressed decay mode $E^- \rightarrow W^- N$ becomes dominant if $M_E > M_N + m_W$ and corresponding decay width is given by\\
\begin{equation}
\Gamma(E^- \rightarrow W^- N) = \frac{M_E}{32\pi}a_E[(1 - r_N^E - r_W^E)^{2} - 4r_N^E r_W^E]^{1/2}[(1 - r_N^E)^{2} + (1 + r_N^E -6\sqrt{r_N^E})r_W^E - 2(r_W^E)^{2}]
\end{equation}
where we neglect terms proportional to the square of mixing angles. Situation becomes more complicated if $M_N < M_E < M_N + m_W$ and needs separate consideration. In the case we are looking at, since N is the lightest VLL, its decays are as in \autoref{5.2}.\\\
Equation (28) transforms to decay width of chiral fourth family (SM4) charged lepton if term $6\sqrt{(r_N^E)}$ is excluded. This term essentially reduces decay width of charged vector-like lepton. For example, when the N and E masses are taken as 500 and 600 GeV, respectively, the decay widths are 0.42 GeV in the VLL case and 1.50 GeV in the SM4 case.
	    \subsection{Heavier neutral lepton}
In this case, additional decay channels are involved, namely, $N \rightarrow W^+ E^-$ if $M_N > M_E + m_W$ and $N \rightarrow E^- l^+ \nu_l, E^- \Bar{q}^{'}q$ if $M_E < M_N < M_E + m_W$ (see Figure 38).\\
\begin{figure}[H]
    \centering
    \includegraphics[width=0.40\textwidth]{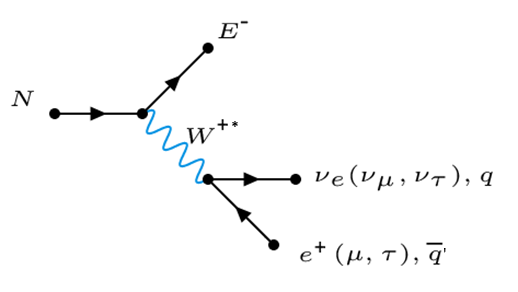}
    \caption{Feynman diagram for heavier neutral lepton decay.}
\end{figure}
\noindent
In iso-singlet case these channels can be neglected since their decay widths are proportional to $(s_L^N s_L^E)^{2}$. Concerning iso-doublet VLLs unsuppressed decay mode $N \rightarrow W^+ E^-$ becomes dominant if $M_N > M_E + m_W$ and corresponding decay width is given by\\
\begin{equation}
\Gamma(N \rightarrow W^+ E^-) = \frac{M_N}{32\pi}a_N[(1 - r_E^N - r_W^N)^{2} - 4r_E^N r_W^N]^{1/2}[(1 - r_E^N)^{2} + (1 + r_E^N -6\sqrt{r_E^N})r_W^N - 2(r_W^N)^{2}]
\end{equation}
where we neglect terms proportional to the square of mixing angles. Situation becomes more complicated if $M_E < M_N < M_E + m_W$ and needs separate consideration. In the case we are looking at, since E is the lightest VLL, its decays are as in \autoref{5.1}.\\\
Equation (29) transforms to decay width of the SM4 neutral lepton if term $6\sqrt{(r_E^N)}$ is excluded. This term essentially reduces decay width of neutral vector-like lepton. For example, when the N and E masses are taken as 600 and 500 GeV, respectively, the decay widths are 0.42 GeV in the VLL case and 1.50 GeV in the SM4 case.
	    \subsection{Long-lived vector-like leptons}
Lightest vector-like leptons may decay in the detector (or even escape it) if mixing angles are extremely small. For example, if $M_N = 500 GeV$ and $s_L^N = 2 \times 10^{-8}$ corresponding path length is approximately 3 m. Neutral vector-like leptons are lost if they escape the detector, whereas charged vector-like leptons still will be seen by the detector.\\\
As mentioned in Introduction, quasi-stable charged vector-like leptons may provide opportunity to realize VLL-catalyzed fusion by analogy with muon-catalyzed fusion if we are lucky. This opportunity, although not very likely, should be explored further. Let us note that nuclear fusion catalyzed by doubly charged scalars have been considered in recent paper \cite{akhmedov2021nuclear}.  
	\section{Process examples}
It is clear that vector-like leptons provide a wide range of opportunities in terms of experimental high energy physics research. Depending on mass and mixing values, they surely provide a lot of different signatures at hadron, lepton, and lepton-hadron colliders. In this aspect, different colliders will be complementary to each other. Below we present a summary of two articles on the subject.
	    \subsection{Associate charged and neutral vector-like leptons production at the LHC}
Associate production of first SM family related vector-like leptons predicted by $E_6$ GUT at the LHC was analyzed in \cite{_zcan_2009}. In this study, charged lepton is assumed to be heavier than the neutral one resulting in the usual decay $E^{\pm} \rightarrow W^{\pm}N$. In the best discovery scenario where the neutral lepton is a Majorana type, the decays of the neutral leptons would yield two electrons (or positrons) fifty-percent of the time: $E^{\pm}N \rightarrow W^{\pm}NN \rightarrow W^{\pm}We^{\pm}We^{\pm}$. When the primary $W^{\pm}$ boson originating from $E^{\pm}$ decays leptonically (e or $\mu$), the resulting third lepton is also of the same sign, yielding in a final state with three same-sign leptons and four jets assuming that secondary W bosons are decay hadronically.  It was shown that the LHC with $\sqrt{s} = 14$ TeV and integrated luminosity 1 $fb^{-1}$ will give opportunity to discover charged vector-like lepton with masses up to $M_E = 250$ GeV. This value can be rescaled \cite{salam2017collider} to 1000 (1900) GeV for integrated luminosity 140 (3000) $fb^{-1}$.
	    \subsection{Pair production of charged vector-like leptons at future colliders}
Recently, pair production of the first generation related charged vector-like leptons at the LHC and future hadron and lepton colliders have been considered in \cite{acar2021search}. It is shown that pair production at the HE-LHC with decay of one of the leptons to Ze and another to He channel, followed by $H \rightarrow b\overline{b}$, and $Z \rightarrow \mu^+ \mu^-$ decays will give opportunity to scan masses of iso-singlet and iso-doublet charged leptons up to 0.9 TeV and 1.5 TeV, respectively. FCC will extend this region up to 2 TeV for iso-singlet and 3.6 TeV for iso-doublet charged leptons.\\\\
At lepton colliders masses of iso-singlet and iso-doublet charged VLL can be scanned up to values close to kinematical limit, $\sqrt{s}/2$. Concerning signature under consideration ILC with $\sqrt{s} = 1$ TeV is comparable with the HL-LHC for iso singlet charged leptons, CLIC or Muon Collider (MC) with $\sqrt{s} = 3$ TeV 
potential exceeds that of HL-LHC both for iso-singlet and iso-doublet and HE-LHC for iso-singlet charged leptons, MC with $\sqrt{s} = 6$ TeV potential exceeds that of HE-LHC both for iso-singlet and iso-doublet and FCC for iso-singlet charged leptons.
	\section{Conclusion}
Vector-like leptons have the same status as vector-like quarks, and there are strong phenomenological arguments supporting the existence of both VLLs and VLQs. Therefore, ATLAS and CMS Collaborations should place more emphasis on investigating charged and neutral vector-like leptons (at least as much as searching for vector-like quarks). Also, the search for VLLs should be included in the physics research programs of future colliders.\\\\
\noindent
There is no reason for the existence of the only one vector-like family. It is much more natural to have vector-like partners for all three SM families. Therefore, one deals with $6\times6$ mass matrices per each SM fermion type ($12\times12$ in neutrino sector if neutral leptons have Majorana masses). It is quiet possible that vector-like leptons and quarks have mass patterns similar to that of SM fermions. In this case vector-like partners of first family (E, N, and D) are the lightest ones among VLLs and VLQs and as a result they will be discovered first. For this reason we focused on them in this paper. However, nobody knows real mass patterns. Therefore, vector-like partners of each SM family may be the lightest ones. Since decay modes in each case are different the search strategy should include each of them, separately.\\\\
\noindent
There is no reason for degenerate mass values of charged and neutral vector-like leptons. It is quite possible that E is essentially heavier than N or vice versa. For this reason, search strategy for charged and neutral VLLs shoud be separated. Moreover, for $M_N \neq M_E$ new decay channels are open for heavier one and this case needs separate analysis as well. If inter-family mixings go in play a lot of different scenarios is occured. In addition some of VLLs may be long-lived. Certainly, VLLs provide a rich phenomenology and require wide range systematic efforts.\\\\
As for the LHC, if $M_N$ and $M_E$ are not much different, associated production looks most promising. From the single production point of view, any significant contribution is not expected from the LHC. If VLLs are not observed at the LHC, then ILC and MC will be out of luck in this regard. However, if the LHC discovers the VLLs, ILC and MC will be enable to get information about mixing angles.\\\\
\noindent
Certainly, the looking for pair and associated productions of vector-like leptons will form the main search strategy. However, single production will also provide important information on VLLs properties if mixing angles are not too small. Since there are many different production and decay channels for charged and neutral vector-like leptons, relevant studies should be done systematically. In particular, the selection of the optimal decay chains for each collider requires detailed signal-background analysis.\\\\
Finally, discovery of VLLs and VLQs will shed light on mass and mixing patterns of fundamental (at today's level) fermions. Mass and mixing patterns of the SM fermions are among the most important issues, which should be clarified in particle physics. In an interview published in CERN Courier \cite{CERNCourier}, Steven Weinberg emphasized this point: “Asked what single mystery, if he could choose, he would like to see solved in his lifetime, Weinberg doesn’t have to think for long: he wants to be able to explain the observed pattern of quark and lepton masses”.
	\section*{Acknowledgement}
We are grateful to Professor Gokhan Unel and Professor Hatice Duran Yildiz for useful discussions.
%    \begin{appendices}
%    \{first}
%    \end{appendices}

\newpage
    \appendix
    \section{Interaction Lagrangians containing mixings}
As a result of including the mixings (Equation (15)) into the Lagrangians (Equations (12)-(14)), we get the following expressions (where notations $\gamma^{-} = \gamma^{\mu}(1-\gamma^{5})$ and $\gamma^{+} = \gamma^{\mu}(1+\gamma^{5})$ are used):\\\
\newline
a) Iso-singlet VLLs

\begin{equation*}
    L_{isosinglet}^{A} = eA_{\mu}\overline{e}\gamma^{\mu}e + eA_{\mu}\overline{E}\gamma^{\mu}E
\end{equation*}

\begin{equation*}
\begin{aligned}
    L_{isosinglet}^{W} 
    & = -\frac{g}{2\sqrt{2}}[W_{\mu}^{-}(c_L^E c_L^N \overline{e}\gamma^- \nu + c_L^E s_L^N \overline{e}\gamma^- N + s_L^E c_L^N \overline{E}\gamma^- \nu + s_L^E s_L^N \overline{E}\gamma^- N)
    \\
    & \phantom{{}={}}+ W_{\mu}^{+}(c_L^E c_L^N \overline{\nu}\gamma^- e + c_L^E s_L^N \overline{N}\gamma^- e + s_L^E c_L^N \overline{\nu}\gamma^- E + s_L^E s_L^N \overline{N}\gamma^- E)]
\end{aligned}
\end{equation*}

\begin{equation*}
\begin{aligned}
    L_{isosinglet}^{Z} 
    & = -\frac{g}{4c_W}Z_{\mu}[(-c_L^E c_L^E + 2s_W^2)\overline{e}\gamma^- e - c_L^E s_L^E \overline{e}\gamma^- E - c_L^E s_L^E \overline{E}\gamma^- e + (-s_L^E s_L^E + 2s_W^2)\overline{E}\gamma^- E
    \\
    & \phantom{{}={}}+ 2s_W^2 \overline{e}\gamma^+ e + 2s_W^2 \overline{E}\gamma^+ E + c_L^N c_L^N \overline{\nu}\gamma^- \nu + c_L^N s_L^N \overline{\nu}\gamma^- N + s_L^N c_L^N \overline{N}\gamma^- \nu + s_L^N s_L^N \overline{N}\gamma^- N]
\end{aligned}
\end{equation*}

\begin{equation*}
\begin{aligned}
    L_{isosinglet}^{H} 
    & = \frac{1}{\eta}c_L^E c_L^E m_e \overline{e}eH + \frac{1}{2\eta}c_L^E s_L^E \overline{e}[(1-\gamma^5)m_e + (1+\gamma^5)m_E]EH
    \\
    & \phantom{{}={}}+ \frac{1}{2\eta}c_L^E s_L^E \overline{E}[(1 - \gamma^5)m_E + (1 + \gamma^5)m_e]eH + \frac{1}{\eta}s_L^E s_L^E m_E \overline{E}EH + \frac{1}{\eta}c_L^N c_L^N m_\nu \overline{\nu}\nu H
    \\
    & \phantom{{}={}}+ \frac{1}{2\eta}c_L^N s_L^N \nu[(1 - \gamma^5)m_\nu + (1 + \gamma^5)m_N]NH
    \\
    & \phantom{{}={}}+ \frac{1}{2\eta}c_L^N s_L^N \overline{N}[(1 - \gamma^5)m_N + (1 + \gamma^5)m_\nu]\nu H + \frac{1}{\eta}s_L^N s_L^N m_N \overline{N}NH
\end{aligned}
\end{equation*}

b) Iso-doublet VLLs

\begin{equation*}
    L_{isodoublet}^{A} = eA_{\mu}\overline{e}\gamma^{\mu}e + eA_{\mu}\overline{E}\gamma^{\mu}E
\end{equation*}

\begin{equation*}
\begin{aligned}
    L_{isodoublet}^{W} 
    & = -\frac{g}{2\sqrt{2}}W_\mu^-[(c_L^E c_L^N + s_L^E s_L^N)\overline{e}\gamma^- \nu + (c_L^E s_L^N - c_L^N s_L^E)\overline{e}\gamma^- N + (c_L^N s_L^E - c_L^E s_L^N)\overline{E}\gamma^- \nu
    \\
    & \phantom{{}={}}+ (s_L^E s_L^N + c_L^E c_L^N)\overline{E}\gamma^- N + s_R^E s_R^N \overline{e}\gamma^+ \nu - c_R^N s_R^E \overline{e}\gamma^+ N - c_R^E s_R^N \overline{E}\gamma^+ \nu + c_R^E c_R^N \overline{E}\gamma^+ N]
    \\
    & \phantom{{}={}}-\frac{g}{2\sqrt{2}}W_\mu^+[(c_L^E c_L^N + s_L^E s_L^N)\overline{\nu}\gamma^- e + (c_L^N s_L^E - c_L^E s_L^N)\overline{\nu}\gamma^- E + (c_L^E s_L^N - c_L^N s_L^E)\overline{N}\gamma^- e
    \\
    & \phantom{{}={}}+ (c_L^E c_L^N + s_L^E s_L^N)\overline{N}\gamma^- E + s_R^N s_R^E \overline{\nu}\gamma^+ e - c_R^E s_R^N \overline{\nu}\gamma^+ E - c_R^N s_R^E\overline{N}\gamma^+ e + c_R^N c_R^E \overline{N}\gamma^+ E]
\end{aligned}
\end{equation*}

\begin{equation*}
\begin{aligned}
    L_{isodoublet}^{Z}
    & = -\frac{g}{4c_W}Z_\mu[(-1 + 2s_W^2)\overline{e}\gamma^- e + (-1 + 2s_W^2)\overline{E}\gamma^- E + (-s_R^E s_R^E + 2s_W^2)\overline{e}\gamma^+ e + c_R^E s_R^E \overline{e}\gamma^+ E
    \\
    & \phantom{{}={}}+ c_R^E s_R^E \overline{E}\gamma^+ e + (-c_R^E c_R^E + 2s_W^2)\overline{E}\gamma^+ E + \overline{\nu}\gamma^- \nu + \overline{N}\gamma^- N + s_R^N s_R^N \overline{\nu}\gamma^+ \nu - c_R^N s_R^N \overline{\nu}\gamma^+ N
    \\
    & \phantom{{}={}}- c_R^N s_R^N \overline{N}\gamma^+ \nu + c_R^N c_R^N \overline{N}\gamma^+ N]
\end{aligned}
\end{equation*}

\begin{equation*}
\begin{aligned}
    L_{isodoublet}^{H} 
    & = \frac{1}{\eta}c_R^E c_R^E m_e \overline{e}eH + \frac{1}{2\eta}c_R^E s_R^E \overline{e}[(1-\gamma^5)m_E + (1+\gamma^5)m_e]EH
    \\
    & \phantom{{}={}}+ \frac{1}{2\eta}c_R^E s_R^E \overline{E}[(1 - \gamma^5)m_e + (1 + \gamma^5)m_E]eH + \frac{1}{\eta}s_R^E s_R^E m_E \overline{E}EH + \frac{1}{\eta}c_R^N c_R^N m_\nu \overline{\nu}\nu H
    \\
    & \phantom{{}={}}+ \frac{1}{2\eta}c_R^N s_R^N \overline{\nu}[(1 - \gamma^5)m_N + (1 + \gamma^5)m_\nu]NH
    \\
    & \phantom{{}={}}+ \frac{1}{2\eta}c_R^N s_R^N \overline{N}[(1 - \gamma^5)m_\nu + (1 + \gamma^5)m_N]\nu H + \frac{1}{\eta}s_R^N s_R^N m_N \overline{N}NH
\end{aligned}
\end{equation*}

    \label{sec:appendix}

	\cleardoublepage
	\phantomsection
	\addcontentsline{toc}{section}{\textbf{References}}
	%\bibliography{ms}
	\vfill
\setlength{\emergencystretch}{14em}
\printbibliography
\end{document}